\documentclass[aps,prd,onecolumn,eqsecnum,amsmath,nofootinbib,preprintnumbers,superscriptaddress]{revtex4-2}

\usepackage[dvipsnames]{xcolor}
\usepackage{color,graphicx,float,subfigure,xcolor}
\usepackage{amsfonts,amssymb,mathrsfs,times}
\usepackage{bm}
\usepackage{multirow}
\usepackage{mathtools}
\usepackage{dsfont}
\usepackage{setspace}
\usepackage{amsmath} 
\usepackage{graphicx}
\usepackage[colorlinks,linkcolor=blue,anchorcolor=blue,citecolor=blue]{hyperref}
\usepackage{ulem}
\usepackage{subcaption}
\usepackage{amsthm}
\newtheorem{theorem}{Theorem}
\usepackage[justification=raggedright,singlelinecheck=false]{caption}

\begin{document}
\title{The structured pseudospectrum of Schwarzschild black hole}
\preprint{\hfill {\small {ICTS-USTC/PCFT-26-79}}}
\date{\today}

\author{Liang-Bi Wu}
\email{wulb@ucas.ac.cn}
\affiliation{School of Fundamental Physics and Mathematical Sciences, Hangzhou Institute for Advanced Study, UCAS, Hangzhou, Zhejiang 310024, China}

\author{Yu-Sen Zhou}
\email{zhou\_ys@mail.ustc.edu.cn}
\affiliation{Interdisciplinary Center for Theoretical Study and Department of Modern Physics,\\
University of Science and Technology of China, Hefei, Anhui 230026, China}

\author{Ming-Fei Ji}
\email{jimingfei@mail.ustc.edu.cn}
\affiliation{Interdisciplinary Center for Theoretical Study and Department of Modern Physics,\\
University of Science and Technology of China, Hefei, Anhui 230026, China}

\author{Xia-Yuan Liu}
\email{liuxiayuan@mail.ustc.edu.cn}
\affiliation{Interdisciplinary Center for Theoretical Study and Department of Modern Physics,\\
University of Science and Technology of China, Hefei, Anhui 230026, China}

\author{Libo Xie}
\email{xielibo23@mails.ucas.ac.cn}
\affiliation{School of Fundamental Physics and Mathematical Sciences, Hangzhou Institute for Advanced Study, UCAS, Hangzhou, Zhejiang 310024, China}
\affiliation{CAS Key Laboratory of Theoretical Physics, Institute of Theoretical Physics, Chinese Academy of Sciences, Beijing 100190, China}
\affiliation{University of Chinese Academy of Sciences, Beijing 100190, China}

\author{Li-Ming Cao}
\email[corresponding author: ]{caolm@ustc.edu.cn}
\affiliation{Interdisciplinary Center for Theoretical Study and Department of Modern Physics,\\
University of Science and Technology of China, Hefei, Anhui 230026, China}
\affiliation{Peng Huanwu Center for Fundamental Theory, Hefei, Anhui 230026, China}


\begin{abstract}
The non-Hermitian nature of black hole quasinormal modes (QNMs) can give rise to remarkable spectrum instability. Such instability is naturally characterized by the pseudospectrum for the operator. However, conventional pseudospectrum analyses permit arbitrary operator perturbations that may not correspond to physically admissible modifications. In this work, we introduce the structured pseudospectrum into black hole perturbation theory. We construct two physically motivated perturbation spaces based on effective field theory and the Rezzolla--Zhidenko  parametrization, and get their structured pseudospectra using two distinct norms. Our numerical results show that imposing physical constraints substantially reduces the accessible spectrum instability, especially for higher overtones. More importantly, at sufficiently small perturbation scales $\varepsilon$, the structured pseudospectrum around isolated QNMs exhibits approximately elliptical contours with preferred directions that depend on both the mode and the perturbation space, revealing an anisotropic structure beyond the familiar circular behavior of the conventional pseudospectrum.

\end{abstract}

\maketitle

\section{Introduction}
When a black hole is perturbed, it relaxes back toward equilibrium through the emission of gravitational waves. In the linear regime, this relaxation is characterized by quasinormal modes (QNMs), whose frequencies form a discrete spectrum of complex-valued eigenvalues. The real part of each QNM gives the oscillation frequency, while the imaginary part governs the damping rate. These modes are fixed by the black hole background and by appropriate boundary conditions, typically purely ingoing waves at the horizon and purely outgoing waves at infinity. As distinctive spectral fingerprints of black holes, QNMs provide a powerful tool for testing the Kerr hypothesis and probing gravitational dynamics in the strong-field regime. This research framework is known as black hole spectroscopy~\cite{Kokkotas:1999bd,Berti:2009kk,Konoplya:2011qq,Berti:2025hly,Cardoso:2025npr}.

Recent advances have increasingly highlighted the non-Hermitian (NH) character of the QNM problem. Since QNMs are defined by purely ingoing boundary conditions at the horizon and purely outgoing boundary conditions at spatial infinity, the resulting spectral problem is inherently non-Hermitian~\cite{Ashida:2020dkc}. NH physics is characterized by the phenomena such as avoided crossings and exceptional points (EPs). In black hole perturbation theory, studies associated with EPs have drawn growing attention in recent years. Avoided crossings (ACs) near EPs can induce resonant excitation between QNMs~\cite{Motohashi:2024fwt,Lo:2025njp,Takahashi:2025uwo,OuldElHadj:2026vym,Jenks:2026tae,Takahashi:2026daf}. If sufficiently many parameters are available, the AC can be tuned to collapse into an EP. It is well known that, for a Kerr black hole where perturbations are described by Teukolsky equation, when quantum numbers $\ell$, $m$ are fixed, the only parameter that can be varied is the reduced angular momentum. Consequently, a Kerr black hole can only exhibit the AC phenomenon~\cite{Motohashi:2024fwt,Lo:2025njp}. Therefore, the mass parameter of scalar~\cite{Cavalcante:2024swt,Cavalcante:2024kmy,Cavalcante:2025abr}, the positive cosmological constant~\cite{Oshita:2025ibu,Rossi:2026als,Nakamoto:2026lyo}, and the electric charge~\cite{Cavalcante:2026vgr,Arcos:2026eis} are considered into the Kerr black hole, and EPs are found accordingly. There are also many studies on EPs in the case of spherically symmetric spacetime. For the case in which a small bump is added to the effective potential, the investigations of EPs are carried out in~\cite{Yang:2025dbn,Cao:2025afs,Hu:2025efp,OuldElHadj:2026vym,OuldElHadj:2026esa,Morikawa:2026fzn,PanossoMacedo:2025xnf}. In addition, there exists the study of the EP for the scalar QNMs of hairy black holes~\cite{Cheng:2026gxu}. Modified QNM boundary conditions can also generate the EPs~\cite{Wu:2025wbp,Long:2026vpw}. The existence of thin spherical shells in spacetime can also lead to EPs~\cite{Wu:2026tfx,Hu:2026tgi}. For higher-dimensional parameter spaces, EPs can form exceptional lines (ELs), which have been explored in gravitational systems~\cite{Cao:2025afs,Nakamoto:2026lyo,Cavalcante:2026vgr}. Such ELs exhibit an intrinsically directional spectral response~\cite{Cao:2025afs,Wu:2026tfx}. By introducing so-called spectral sensitivity anisotropy matrix, we can quantitatively investigate the spectral splitting anisotropy associated with the EL~\cite{Wu:2026tfx}. In the time domain, the usual expansion of the ringdown signal as a sum of independent QNMs can fail in the vicinity of an EP. For a second-order EP, a contribution proportional to $t\mathrm{e}^{-\mathrm{i}\omega t}$ appears~\cite{Yang:2025dbn,PanossoMacedo:2025xnf,Cheng:2026gxu,Imafuku:2026rpn}.

NH physics is characterized not only by phenomena such as ACs and EPs but also by spectrum instability. In the field of black hole perturbations, the history of studying spectrum instability is even longer. Such question was first raised in~\cite{Nollert:1996rf}, where the segmented approximation of effective potential is studied. One can refer to similar approaches for studying spectrum instability in~\cite{Qian:2020cnz,Daghigh:2020jyk,Xie:2025jbr,Wu:2025sbq,Shen:2026hdl}. A small smooth correction to the effective potential is also a widely studied model for investigating spectrum instability~\cite{Berti:2022xfj,Cheung:2021bol,Courty:2023rxk,Yang:2024vor,Cardoso:2024mrw,Ianniccari:2024ysv,MalatoCorrea:2025iuc,Shen:2025nsl,Hu:2025efp,Jia:2026vnx}. A negative deformation of the potential may give rise to dynamical instability of spacetime~\cite{Mai:2025mim,Ma:2026eaf,Cao:2024sot}. Recently, Ref. \cite{DellaRocca:2026zym} indicates that although a secondary bump in the effective potential generated by matter fields around black holes may lead to spectrum instabilities, in anisotropic-fluid models it corresponds to unphysical matter configurations. Small perturbations to the QNM boundary conditions are capable of inducing appreciable shifts in the spectra as well~\cite{Oshita:2025ibu,Solidoro:2024yxi,Destounis:2025dck,Wu:2025wbp,Xie:2026yjg}.

Known alterations to the effective potential enable us to ascertain precisely how the spectrum migrates as the parameters are varied. Nevertheless, if only the strength of the operator modification is known, then although the precise spectrum displacement cannot be ascertained, the range of spectrum migration can still be inferred by computing the pseudospectrum. The pseudospectrum is a norm-dependent diagnostic. It quantifies the smallest perturbation, within a specified perturbation class and norm, that can make a trial frequency an eigenvalue of a perturbed operator. The pseudospectrum approach to the (in)stability of black hole QNMs was first studied in~\cite{Jaramillo:2020tuu,Jaramillo:2022kuv,Gasperin:2021kfv}. Building on this work, a substantial body of work concerning pseudospectra has emerged~\cite{Destounis:2021lum,Cao:2024oud,Cao:2024sot,Arean:2024afl,Garcia-Farina:2024pdd,Arean:2023ejh,Boyanov:2023qqf,Cownden:2023dam,Carballo:2025ajx,Sarkar:2023rhp,Destounis:2023nmb,Luo:2024dxl,Warnick:2024usx,Chen:2024mon,Boyanov:2022ark,Siqueira:2025lww,Besson:2024adi,dePaula:2025fqt,Cai:2025irl,Cao:2025qws,Zhou:2025xdo,Zhang:2026cjq,Ferreira:2026csr,Jia:2026eqw}. More recently, the pseudospectrum behavior near the EPs has also been studied~\cite{Cao:2025afs,Destounis:2026qne}.

Localized deformations of the effective potential have been shown to destabilize even the fundamental QNMs~\cite{Cheung:2021bol}, whereas subsequent investigations have emphasized that some configurations responsible for such instabilities may be incompatible with physically acceptable matter models~\cite{Cardoso:2024mrw,DellaRocca:2026zym}. The central question is therefore not merely whether the QNM spectrum is sensitive to small perturbations, but which aspects of this instability persist when the perturbations are restricted to a specified physical model. More generally, the structured pseudospectrum (see Chapter $50$ in~\cite{trefethen2020spectra}) provide a framework for incorporating prescribed perturbation constraints into spectrum stability analysis. The physical significance of the structured pseudospectrum is that it quantifies, within a prescribed class of physically admissible perturbations, the instability of black hole QNM spectra and the range over which they may shift. This motivates a model-dependent assessment of spectrum stability, in which both the admissible family of deformations and the norm used to quantify their magnitude are specified explicitly. In this work, taking the Schwarzschild black hole as reliable model, we apply the structured pseudospectrum to black hole perturbation theory for the first time.

The remainder of this work is organized as follows. In Sec. \ref{sec: perturbation_space_constructions}, we show the constructions of physical perturbation spaces. In Sec. \ref{sec: perturbation_space_constructions}, we give the definitions of the structured pseudospectrum within two different norms. In Sec. \ref{sec: results_pseudospectrum}, the numerical results of the structured pseudospectrum are shown. Sec. \ref{sec: conclusions} is the conclusions and discussion. In Appendix \ref{app: construction_S}, we introduce physically motivated construction of the structured perturbation space. In Appendix \ref{app: structured_distance}, we transform the solving of the structured pseudospectrum into the optimization problem. Appendix \ref{app: inequality_between_three_d_proof} presents the chain of inequalities relating the three spectral distances including one unstructured spectral distance and two structured spectral distances. Appendix \ref{app: elliptical_proof} gives a proof of the local elliptical asymptotics of the pseudospectrum contours for $d_{\mathrm{str},\mathcal{S}}(\omega)$.

\section{Physical constructions of perturbation spaces}\label{sec: perturbation_space_constructions}
In the Schwarzschild black hole, the perturbation equation is given by
\begin{eqnarray}\label{master_perturbation_equation}
    \Big[\frac{\partial^2}{\partial t^2}-\frac{\partial^2}{\partial x^2}+V(r(x))\Big]\Psi(t,x)=0\, ,
\end{eqnarray}
where the effective potential $V(r)$ is written as
\begin{eqnarray}\label{effective_potential}
    V(r)=\Big(1-\frac{2M}{r}\Big)\Big[\frac{\ell(\ell+1)}{r^2}+\frac{2M(1-s^2)}{r^3}\Big]\, ,
\end{eqnarray}
where $M$ is the mass of Schwarzschild black hole. The compact hyperboloidal coordinates $(\tau, \sigma, \theta, \varphi)$ related with original coordinate $(t, r, \theta, \varphi)$ are defined as follows
\begin{eqnarray}\label{compact_hyperboloidal_coordinates}
    t&=&\tau-h(\sigma)\, ,\nonumber\\
    r&=&\frac{r_{\text{h}}}{\sigma}\, ,
\end{eqnarray}
where $h(\sigma)$ is the height function and $r_{\text{h}}$ denotes the event horizon of Schwarzschild black hole. The coordinate $\sigma$ spans the interval $[0,1]$, with $\sigma=0$ corresponding to null infinity $\mathscr{I}^+$ and $\sigma=1$ to the event horizon $\mathscr{H}^+$. Here, we adopt the height function~\cite{Jaramillo:2020tuu}
\begin{eqnarray}
    h(\sigma)=2M\Big[-\frac{1}{\sigma}+\ln\sigma+\ln(1-\sigma)\Big]\, ,
\end{eqnarray}
and set $M=1$, i.e., $r_{\text{h}}=2$ for the unit. In these coordinates, consider the Fourier transform of $U(\tau,\sigma)$ with respect to time $\tau$
\begin{eqnarray}
    U(\tau, \sigma)=\mathrm{e}^{\mathrm{i}\omega\tau}U(\sigma)\, ,
\end{eqnarray} the master perturbation equation transforms to an eigenvalue problem
\begin{eqnarray}\label{maineq}
    (\omega I-L)U=0\, ,\quad
    L=\frac{1}{\mathrm{i}}
    \begin{bmatrix}
        0   & 1   \\
        L_1 & L_2
    \end{bmatrix}\, ,\quad
    U=\begin{bmatrix}
        \Psi \\
        \Pi
    \end{bmatrix}\, ,
\end{eqnarray}
with $\Pi\equiv\partial_{\tau}\Psi$. The differential operators $L_1$ and $L_2$ are given by
\begin{eqnarray}
    L_{1}(\sigma)&=&\frac{1}{w(\sigma)}\Big[\partial_{\sigma}(p(\sigma)\partial_{\sigma})-q(\sigma)\Big]\, ,\label{operator_L1}\\
    L_{2}(\sigma)&=&\frac{1}{w(\sigma)}\Big[2\gamma(\sigma)\partial_{\sigma}+\partial_{\sigma}\gamma(\sigma)\Big]\, ,\label{operator_L2}
\end{eqnarray}
where above functions read
\begin{eqnarray}\label{function_p_gamma_w_qx}
    p(\sigma)=\frac{\sigma^2f(r(\sigma))}{r_{\text{h}}}\, ,\quad \gamma(\sigma)=h^{\prime}(\sigma)p(\sigma)\, ,\quad w(\sigma)=\frac{1-\gamma^2(\sigma)}{p(\sigma)}\, ,\quad q(\sigma)=\frac{V(r(\sigma))}{p(\sigma)}\, ,
\end{eqnarray}
with $f(r)=1-2M/r$, and the prime denotes differentiation with respect to $\sigma$. With the convention $U\propto\mathrm{e}^{\mathrm{i}\omega\tau}$ adopted in this work, the damped QNMs lie in the upper half of the complex-frequency plane.

The non-Hermitian (NH) nature of the QNM problem implies that its spectrum can be highly sensitive to small perturbations. Previous pseudospectrum analyses have revealed significant spectrum instabilities of black hole QNM overtones~\cite{Jaramillo:2020tuu,Destounis:2021lum,Gasperin:2021kfv,Boyanov:2022ark,Boyanov:2023qqf,Cai:2025irl,Zhang:2026cjq}. However, the unstructured (ordinary) pseudospectrum allows arbitrary operator perturbations, many of which may not correspond to physically realizable modifications of the underlying black hole perturbation equations. In realistic physical models, the admissible perturbations are often subject to additional constraints. Consequently, the unstructured pseudospectrum may overestimate the spectrum instability associated with a prescribed class of physical perturbations. This motivates the study of the structured pseudospectrum, which restricts perturbations to a physically admissible space and thereby provides a more physically informative characterization of QNM spectrum stability.

A widely studied model is to change the effective potential to investigate the stability of QNM spectra, which is originated from~\cite{Cheung:2021bol}. Motivated by it, physically, we generally only allow perturbations of the effective potential, namely
\begin{eqnarray}\label{V_to_V_add_deltaV}
    V(\sigma)\to V(\sigma)+\delta V(\sigma)\, ,\quad \delta V\in\mathcal{S}\, ,
\end{eqnarray}
where $\mathcal{S}$ is some function space, and $\delta V$ is a real function of $\sigma$. The perturbation $\delta L$ of the operator $L$ induced by $\delta V$ reads
\begin{eqnarray}\label{deltaL_deltaV}
    \delta L[\delta V]=-\frac{1}{\mathrm{i}}
    \begin{bmatrix}
        0 & 0\\
        \dfrac{\delta V(\sigma)}
        {p(\sigma)w(\sigma)}& 0
    \end{bmatrix}\, .
\end{eqnarray}
Thus, the physically relevant quantity is not only the function $\delta V(\sigma)$ itself, but also the perturbation $\delta L[\delta V]$ it induces in the operator $L$. We restrict the perturbation of potential to a finite-dimensional admissible function space which denoted by
\begin{eqnarray}\label{Space_S}
    \mathcal S:=\operatorname{span}\Big\{\phi_1(\sigma),\phi_2(\sigma),\ldots,\phi_m(\sigma)\Big\}\, ,
\end{eqnarray}
where $\{\phi_k\}_{k=0}^m$ are assumed to be real-valued and linearly independent. In the followings, we call $\mathcal{S}$ the structured perturbation space. Since $\delta V\in \mathcal{S}$, every admissible perturbation admits the following expansion
\begin{eqnarray}\label{deltaV_expansion}
    \delta V(\sigma)=\sum_{k=1}^{m}c_k\phi_k(\sigma)\, ,\quad c_k\in\mathbb{R}\, .
\end{eqnarray}
Physical constructions of the perturbation space $\mathcal{S}$ are shown in Appendix \ref{app: construction_S}. In the present work, we give two (physical) structured perturbation spaces, which are written as $\mathcal{S}_{\mathrm{EFT}}$ and $\mathcal{S}_{\mathrm{RZ}}$, respectively.

\section{Structured pseudospectrum}
In Sec. \ref{sec: perturbation_space_constructions}, we present the allowed space for the effective potential perturbation $\delta V$. In this section, we will define the structured $\varepsilon$-pseudospectrum based on $\mathcal{S}$ and the original operator $L$. First of all, we give a brief review on the unstructured pseudospectrum~\cite{Jaramillo:2020tuu} of the Schwarzschild black hole for comparison with the structured pseudospectrum. Unlike the structured perturbations considered in the main text, the unstructured (ordinary) pseudospectrum allows arbitrary operator perturbations without restricting them to those induced by potential deformations $\delta V\in \mathcal{S}$.

Consider the hyperboloidal first-order formulation with time $\tau$
\begin{eqnarray}
    \partial_{\tau} U = \mathrm{i} L U,
    \qquad
    U=
    \begin{bmatrix}
        \Psi\\
        \Pi
    \end{bmatrix}\, ,\quad \text{with}\quad L=
    \frac{1}{\mathrm{i}}
    \begin{bmatrix}
        0 & 1\\
        L_1 & L_2
    \end{bmatrix}\, .
\end{eqnarray}
The energy norm $\|\cdot\|_{\mathrm{E}}$ associated with the master variable $U$ can be written as~\cite{Jaramillo:2020tuu,Gasperin:2021kfv,Jaramillo:2022kuv,Besson:2024adi}
\begin{eqnarray}\label{eq:appB_energy}
    E[U]=\frac{1}{2}\int_0^1\Big[p(\sigma)|\partial_{\sigma}\Psi|^2+q(\sigma)|\Psi|^2+w(\sigma)|\Pi|^2\Big]\mathrm{d}\sigma\, ,\quad \|U\|_\mathrm{E}=\sqrt{E[U]}\, .
\end{eqnarray}
This motivates the following energy inner product
\begin{eqnarray}\label{eq:appB_energy_inner}
    \left\langle U_1,U_2\right\rangle_{\mathrm{E}}=\frac{1}{2}\int_0^1\Big[p(\sigma)\partial_{\sigma}\overline{\Psi}_1\partial_{\sigma}\Psi_2+q(\sigma)\overline{\Psi}_1\Psi_2+w(\sigma)\overline{\Pi}_1\Pi_2
    \Big]
    \mathrm{d}\sigma\, .
\end{eqnarray}
where a bar represents the complex conjugate. Note that Eq. \eqref{eq:appB_energy_inner} is determined by the dynamics of the black hole perturbation theory itself. It is therefore more appropriate for measuring physically relevant perturbations than the $L^2$-norm.

To make the problem more transparent, we consider the matrix version. After some discretization such as using Chebyshev--Lobatto (CL) grid, inner product (\ref{eq:appB_energy_inner}) can be written as\footnote{We use bold symbols to represent matrices and vectors which are the discrete versions of operator and field.} 
\begin{eqnarray}\label{inner_product_discretization}
    \langle \mathbf{U}_1,\mathbf{U}_2\rangle_{\mathrm{E}}=\mathbf{U}_1^\dagger \mathbf{H}_{\mathrm{E}} \mathbf{U}_2\, ,
\end{eqnarray}
where $(\cdot)^{\dagger}$ stands for the Hermitian conjugation. Since for Schwarzschild black hole, three functions $p(\sigma)$, $q(\sigma)$ and $w(\sigma)$ are all positive on $[0,1]$, the matrix $\mathbf{H}_{\mathrm{E}}$ is a positive definite Hermitian matrix. Therefore, it admits a Cholesky decomposition, i.e., $\mathbf{H}_{\mathrm{E}}=\mathbf{W}^{\dagger}\mathbf{W}$, where $\mathbf{W}$ is an upper triangular matrix. The matrix $\mathbf{W}$ whitens the energy in the sense that the Euclidean norm ($2$-norm) of the energy-normalized state $\widetilde{\mathbf{U}} = \mathbf{W}\mathbf{U}$ equals the physical energy norm of $\mathbf{U}$, i.e.,
\begin{eqnarray}
    \|\widetilde{\mathbf{U}}\|_2^2=\widetilde{\mathbf{U}}^\dagger\widetilde{\mathbf{U}}=\mathbf{U}^\dagger \mathbf{H}_{\mathrm{E}}\mathbf{U}=\|\mathbf{U}\|_{\mathrm{E}}^2\, .
\end{eqnarray}
Here, $\|\cdot\|_2$ denotes the Euclidean norm for vectors. Correspondingly, the counterpart of the perturbed operator acting on energy normalized states is 
\begin{eqnarray}\label{tilde_deltaL}
    \widetilde{\delta\mathbf{L}}=\mathbf{W}(\delta \mathbf{L})\mathbf{W}^{-1}\, .
\end{eqnarray}

We adopt the energy inner product defined in Eq. (\ref{eq:appB_energy_inner}), which induces the energy norm $\|U\|_\mathrm{E}=\sqrt{E[U]}$. Afterwards, the corresponding induced operator norm is defined as
\begin{eqnarray}\label{induced_operator_norm}
    \|A\|_\mathrm{E}:=\sup_{U\neq 0}\frac{\|AU\|_{\mathrm{E}}}{\|U\|_{\mathrm{E}}}\, .
\end{eqnarray}
After discretization, the energy inner product can be written as Eq. (\ref{inner_product_discretization}). Using the Cholesky decomposition, and introducing the energy-normalized state $\widetilde{\mathbf{U}}=\mathbf{W} \mathbf{U}$, we obtain
\begin{eqnarray}\label{spectral_norm_matrix}
    \|\mathbf{A}\|_{\mathrm{E}}=\sup_{\widetilde{\mathbf{U}}\neq 0}\frac{\|\mathbf{W}\mathbf{A}\mathbf{W}^{-1}\widetilde{\mathbf{U}}\|_2}{\|\widetilde{\mathbf{U}}\|_2}=\|\mathbf{W}\mathbf{A}\mathbf{W}^{-1}\|_2\, .
\end{eqnarray}
Here, $\|\cdot\|_2$ denotes the induced spectral norm for matrices, which equals to the maximum singular value of the matrix. The unstructured $\varepsilon$-pseudospectrum $\sigma_{\varepsilon}^{\rm unstr}(L)$ of a operator $L$ is defined as
\begin{eqnarray}\label{unstructured_eps_pseudospectrum}
    \sigma_{\varepsilon}^{\rm unstr}(L)=\Big\{\omega\in\mathbb C\Big|\ \exists\,\delta L,\,\|\delta L\|_{\mathrm{E}}\leq\varepsilon,\,0\in\lambda(\omega I-L-\delta L)\Big\}\, ,
\end{eqnarray}
where $\delta L$ is an arbitrary complex operator perturbation and $\lambda(\cdot)$ stands for the eigenvalue. For convenience, ``unstr'' stands for unstructured. For a fixed frequency $\omega\notin\lambda(L)$, the corresponding unstructured spectral distance $d_{\rm unstr}(\omega)$ is defined as
\begin{eqnarray}\label{unstr_d}
    d_{\rm unstr}(\omega):=\inf_{\delta L}\Big\{\|\delta L\|_{\mathrm{E}}\ \Big| \; 0\in\lambda(\omega I-L-\delta L)\Big\}\, .
\end{eqnarray}
This quantity measures the minimum energy-norm operator perturbation required to make $\omega$ an eigenvalue of the perturbed operator. Accordingly, the unstructured $\varepsilon$-pseudospectrum can be expressed as
\begin{eqnarray}\label{unstr_sigma}
    \sigma_{\varepsilon}^{\rm unstr}(L)=\Big\{\omega\in\mathbb C\;\Big|\; d_{\rm unstr}(\omega)\leq\varepsilon\Big\}\, .
\end{eqnarray}
In other words, the pseudospectrum is the level set of the spectral distance. Mathematically, it can be proven that the unstructured spectral distance for the matrix version can be obtained as follow
\begin{eqnarray}\label{unstr_d_matrix}
    d_{\rm unstr}(\omega)=\sigma_{\min}\Big[\mathbf{W}(\omega \mathbf{I}-\mathbf{L})\mathbf{W}^{-1}\Big]\, ,
\end{eqnarray}
in which $\sigma_{\min}(\cdot)$ is the minimum singular value. Therefore, given $\omega\in\mathbb{C}$, one can evaluate Eq. (\ref{unstr_d_matrix}) to get the unstructured pseudospectrum.

Based on the definition of unstructured pseudospectrum (\ref{unstructured_eps_pseudospectrum}), from Eq. (\ref{deltaL_deltaV}), it is natural to derive the definition of the structured $\varepsilon$-pseudospectrum by requiring $\delta V\in\mathcal{S}$. This means that
\begin{eqnarray}\label{structured_eps_pseudospectrum_E}
    \sigma_\varepsilon^{\mathrm{str,E}}(L)=\Big\{\omega\in\mathbb C\;\Big|\;\exists\,\delta V\in \mathcal{S},\;\|\delta L[\delta V]\|_{\mathrm{E}}\le\varepsilon,\;0\in\lambda(\omega I-L-\delta L)\Big\}\, .
\end{eqnarray}
The structured perturbation of the operator $\delta L[\delta V]$, which is induced by the expansion of the perturbation of the potential (\ref{deltaV_expansion}), has the following explicit form
\begin{eqnarray}\label{delta_L_in_terms_of_phi}
    \delta L[\delta V](\sigma)=-\frac{1}{\mathrm{i}}\sum_{k=1}^{m}c_k M_k(\sigma)\, ,\quad \text{with} \quad M_k(\sigma)=\begin{bmatrix}
    0&0\\
    \dfrac{\phi_k(\sigma)}
    {p(\sigma)w(\sigma)}&0
    \end{bmatrix}\, ,
\end{eqnarray}
where the lower-left entry is understood as a multiplication operator. For a fixed frequency $\omega\notin\lambda(L)$, the corresponding structured spectral distance is 
\begin{eqnarray}\label{structured_spectral_distance_E}
    d_{\mathrm{str},\mathrm{E}}(\omega)=\inf_{\delta V\in\mathcal{S}}\Big\{\|\delta L[\delta V]\|_{\mathrm{E}}\;\Big|\;0\in\lambda(\omega I-L-\delta L)\Big\}\, .
\end{eqnarray}
Accordingly, the structured $\varepsilon$-pseudospectrum is
\begin{eqnarray}\label{str_sigma_E}
    \sigma_{\varepsilon}^{\mathrm{str},\mathrm{E}}(L)=\Big\{\omega\in\mathbb C\;\Big|\; d_{\mathrm{str},\mathrm{E}}(\omega)\leq\varepsilon\Big\}\, .
\end{eqnarray}

In the definition of $\sigma_\varepsilon^{\mathrm{str,E}}(L)$ in Eq. (\ref{structured_eps_pseudospectrum_E}), the size of the perturbation is measured in the energy norm $\|\cdot\|_{\mathrm{E}}$. This choice is physically motivated, but it is not essential, and no inner product on $\mathcal{S}$ is needed for that definition. Other norms can be used equally well. For instance, one may equip $\mathcal{S}$ with an inner product $\langle\cdot,\cdot\rangle_{\mathcal{S}}$ and measure $\delta V$ by the induced norm $\|\delta V\|_{\mathcal{S}} = \sqrt{\langle\delta V,\delta V\rangle_{\mathcal{S}}}$. From Eq. (\ref{tilde_deltaL}), we therefore define the inner product on $\mathcal{S}$ by pulling back the Hilbert--Schmidt (HS) inner product of the corresponding energy normalized operator perturbations,
\begin{eqnarray}\label{inner_product_S}
    \left\langle\delta V_1,\delta V_2\right\rangle_{\mathcal{S}}:=\operatorname{Re}\operatorname{Tr}\Big[(\widetilde{\delta\mathbf{L}}_1)^{\dagger}(\widetilde{\delta\mathbf{L}}_2)\Big]=\operatorname{Re}\operatorname{Tr}\Bigg[\Big(\mathbf{W}\delta \mathbf{L}[\delta V_1]\mathbf{W}^{-1}\Big)^\dagger\Big(\mathbf{W}\delta \mathbf{L}[\delta V_2]\mathbf{W}^{-1}\Big)\Bigg]\, ,\quad \delta V_1\, ,\delta V_2\in \mathcal{S}\, ,
\end{eqnarray}
where the formulation $\|\mathbf{A}\|_{\mathrm{HS}}^2\equiv\operatorname{Tr}(\mathbf{A}^{\dagger}\mathbf{A})$ is called the Hilbert--Schmidt (HS) norm or the Frobenius norm (namely the square root of the sum of the squared moduli of all matrix elements, and $\operatorname{Tr}$ is the trace. Mathematically, it is a pullback inner product from the space of operators to the admissible potential space $\mathcal{S}$. Since the admissible potential perturbations and their expansion coefficients are real, $\mathcal{S}$ is regarded as a real vector space. For the present Schwarzschild problem with real potential perturbations, the resulting trace is real analytically, so the operation $\operatorname{Re}$ is formally redundant but is retained to make the real structure explicit and to suppress spurious imaginary parts arising from numerical roundoff.

Since $\delta V$ has the expansion on the space $\mathcal{S}$ [see Eq. (\ref{deltaV_expansion})], we therefore construct a real symmetric Gram matrix associated with the basis
$\{\phi_k\}_{k=1}^{m}$ which reads
\begin{eqnarray}\label{Gram_definition}
    \mathbf{G}_{ij}:=\langle\phi_i,\phi_j\rangle_{\mathcal{S}}=\operatorname{Re}\operatorname{Tr}\Bigg[\Big(\delta \mathbf{L}[\phi_i]\Big)^{\dagger}\mathbf{H}_{\mathrm{E}}\Big(\delta \mathbf{L}[\phi_j]\Big)\mathbf{H}_{\mathrm{E}}^{-1}\Bigg]\, ,
\end{eqnarray}
where the second equality is obtained by the cyclic property of the trace. Combining Eq. (\ref{Gram_definition}) and Eq. (\ref{deltaV_expansion}), it is easy to derive that $\|\delta V\|^2_{\mathcal{S}}$ can indeed be written in a real quadratic form, which means that
\begin{eqnarray}\label{delta_V_norm}
    \|\delta V\|_{\mathcal{S}}=\sqrt{\mathbf{c}^{\mathrm{T}}\mathbf{G}\mathbf{c}}=\|\widetilde{\delta \mathbf{L}}\|_{\mathrm{HS}}\, ,
\end{eqnarray}
in which $(\cdot)^\mathrm{T}$ corresponds to the transpose. Based on the above norm of $\delta V$ in $\mathcal{S}$, another structured $\varepsilon$-pseudospectrum of $L$ is definable, which is defined as
\begin{eqnarray}\label{structured_eps_pseudospectrum_S}
    \sigma_\varepsilon^{\mathrm{str},\mathcal{S}}(L)=\Big\{\omega\in\mathbb C\;\Big|\;\exists\,\delta V\in \mathcal{S},\;\|\delta V\|_{\mathcal{S}}\le\varepsilon,\;0\in\lambda(\omega I-L-\delta L)\Big\}\, .
\end{eqnarray}
For a fixed frequency $\omega\notin\lambda(L)$, the corresponding‌ structured spectral distance is given by
\begin{eqnarray}\label{structured_spectral_distance_S}
    d_{\mathrm{str},\mathcal{S}}(\omega)=\inf_{\delta V\in\mathcal{S}}\Big\{\|\delta V\|_{\mathcal{S}}\;\Big|\;0\in\lambda(\omega I-L-\delta L)\Big\}\, .
\end{eqnarray}
Consequently, the structured $\varepsilon$-pseudospectrum of $L$ associated with the inner product of $\mathcal{S}$ is
\begin{eqnarray}
    \sigma_\varepsilon^{\mathrm{str},\mathcal{S}}(L)=\Big\{\omega\in\mathbb C\;\Big|\;d_{\mathrm{str},\mathcal{S}}(\omega)\le\varepsilon\Big\}\, .
\end{eqnarray}

In summary, two notions of structured spectral distances have been introduced [Eq. (\ref{structured_spectral_distance_E}) and Eq. (\ref{structured_spectral_distance_S})]. As for the methods of calculating these two structured spectral distances, one can refer to Appendix \ref{app: structured_distance}. More importantly, for a fixed $\omega$, we are able to prove the following inequality between previous mentioned three distances $d_{\rm unstr}(\omega)$, $d_{\mathrm{str},\mathrm{E}}(\omega)$, and $d_{\mathrm{str},\mathcal{S}}(\omega)$, namely
\begin{eqnarray}\label{inequality_between_three_d}
    0\leq d_{\rm unstr}(\omega) \leq d_{\mathrm{str},\mathrm{E}}(\omega) \leq d_{\mathrm{str},\mathcal{S}}(\omega) \leq \sqrt{N_{\text{c}}} d_{\mathrm{str},\mathrm{E}}(\omega)\, ,
\end{eqnarray}
where $N_{\text{c}}$ is the numerical resolution of the operator from which the dimension of the matrix $\mathbf{L}$ is $2(N_{\text{c}}+1)$. For a more detailed proof of (\ref{inequality_between_three_d}), one can refer to Appendix \ref{app: inequality_between_three_d_proof}.

\section{Results of structured pseudospectra}\label{sec: results_pseudospectrum}
In this section, we compare the unstructured pseudospectrum with the structured pseudospectra obtained by restricting the perturbation of the effective potential to the physically motivated spaces introduced in Sec. \ref{sec: perturbation_space_constructions}. In the present work, we consider the EFT-matched perturbation space $\mathcal{S}_{\mathrm{EFT}}^{(\mathcal{N})}$ and the RZ-matched perturbation space $\mathcal{S}_{\mathrm{RZ}}^{(\mathcal{N})}$, whose explicit constructions are given in Eqs. (\ref{S_EFT}) and (\ref{S_RZ}) of Appendix \ref{app: construction_S}, respectively. Two structured spectral distances $d_{\mathrm{str},\mathrm{E}}(\omega)$ and $d_{\mathrm{str},\mathcal{S}}(\omega)$ are defined in Eqs. (\ref{structured_spectral_distance_E}) and (\ref{structured_spectral_distance_S}). Their numerical evaluations are formulated as the optimization problems described in Appendix \ref{app: structured_distance}. For comparison, the unstructured spectral distance $d_{\rm unstr}(\omega)$ is directly evaluated from Eq. (\ref{unstr_d_matrix}). Once a spectral distance $d(\omega)$ is obtained over the complex-frequency plane, the corresponding $\varepsilon$-pseudospectrum is its level set with $d(\omega)\leq\varepsilon$. According to Eq. \eqref{inequality_between_three_d}, the corresponding pseudospectra satisfy
\begin{eqnarray}\label{pseudospectrum_inclusion}
    \sigma_{\varepsilon/\sqrt{N_{\text{c}}}}^{\mathrm{str},\mathrm{E}}(\mathbf{L}) \subseteq \sigma_{\varepsilon}^{\mathrm{str},\mathcal{S}}(\mathbf{L}) \subseteq \sigma_{\varepsilon}^{\mathrm{str},\mathrm{E}}(\mathbf{L}) \subseteq \sigma_{\varepsilon}^{\rm unstr}(\mathbf{L})\, .
\end{eqnarray}
Therefore, for a fixed perturbation scale $\epsilon$, the unstructured pseudospectrum contains the largest accessible spectral region, while the two structured pseudospectra are further restricted by the prescribed perturbation space and by the norm used to measure the perturbation.

\begin{figure}[htbp]
    \centering
    \subfigure[]
    {\includegraphics[width=0.47\linewidth]{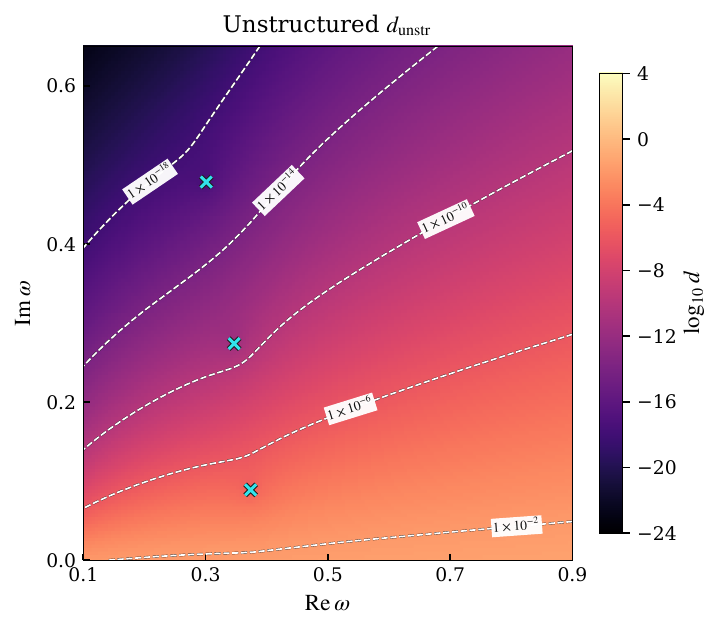}\label{unstr_d_EFT}}\hfill
    \subfigure[]
    {\includegraphics[width=0.47\linewidth]{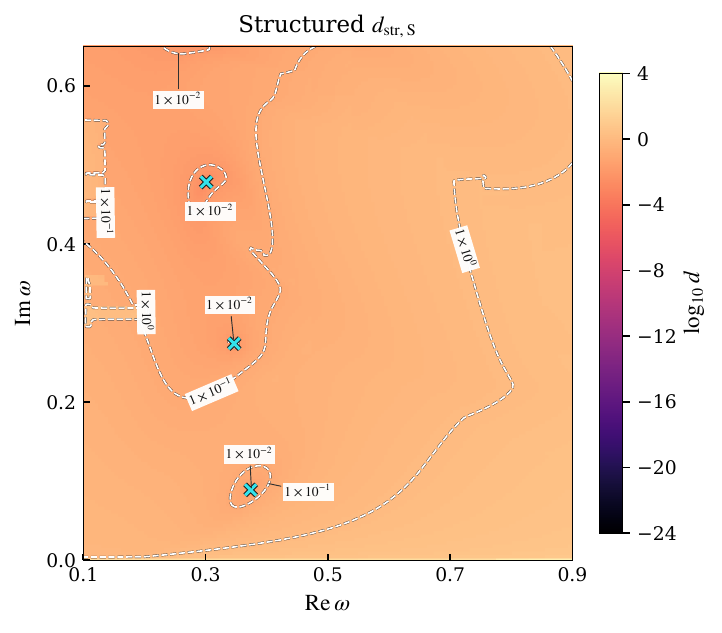}\label{str_d_S_EFT}}\\
    \subfigure[]
    {\includegraphics[width=0.47\linewidth]{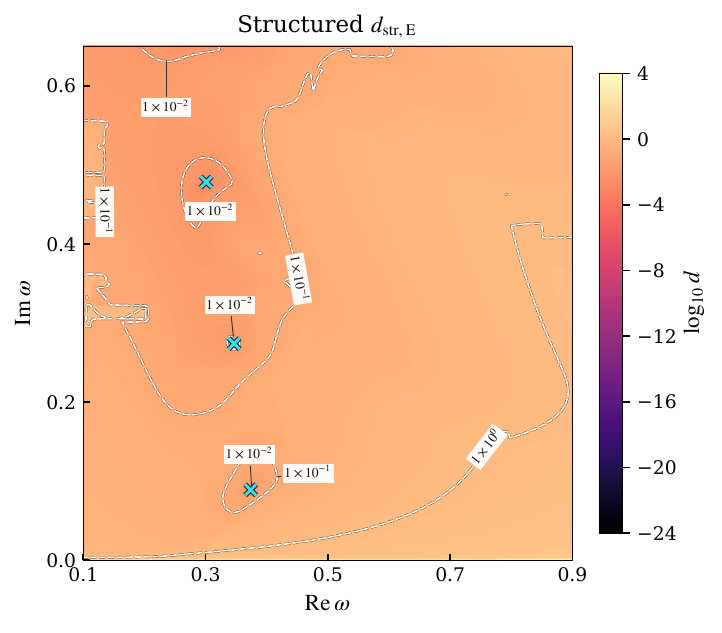}\label{str_d_E_EFT}}\hfill
    \subfigure[]
    {\includegraphics[width=0.47\linewidth]{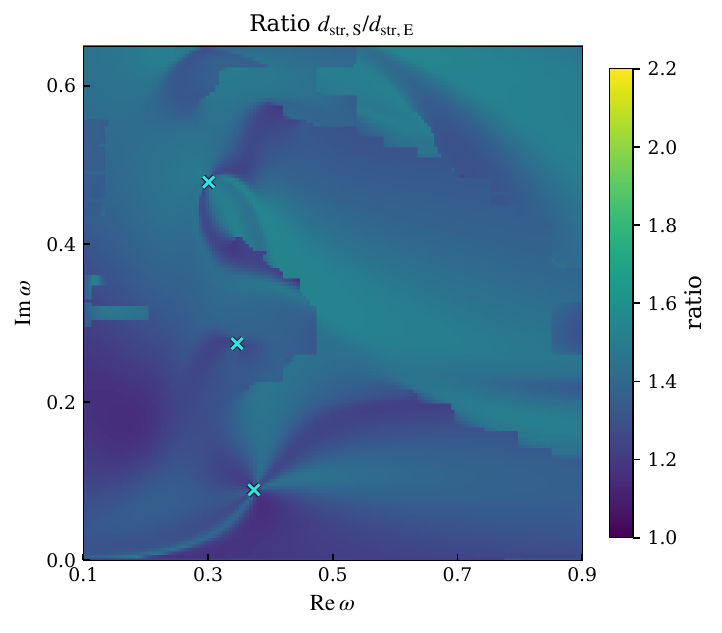}\label{Ratio_d_EFT}}
    \caption{Unstructured and EFT-structured pseudospectra of the Schwarzschild black hole for axial gravitational perturbations $(s=2)$ with $m=4$. Panels (a)-(c) show the unstructured pseudospectrum $d_{\rm unstr}(\omega)$, the structured pseudospectrum $d_{\mathrm{str},\mathcal{S}}(\omega)$ defined with the HS norm, and $d_{\mathrm{str},\mathrm{E}}(\omega)$ that defined with the energy norm, respectively. These three panels use the same color legend for $\log_{10}d$, allowing a direct comparison of the spectral distances. The structured spectral distances are substantially larger than the unstructured one and vary more slowly over the complex-frequency plane, with noticeably smaller gradients. The dashed curves are selected $\varepsilon$-pseudospectrum contours. The crosses mark the original QNMs. Panel (d) shows the ratio $d_{\mathrm{str},\mathcal{S}}/d_{\mathrm{str},\mathrm{E}}$. The numerical set up is that the resolution $N_{\text{c}}=120$, and the frequency grid is $150\times150$.}
    \label{fig:EFT_global_pseudospectrum}
\end{figure}

\begin{figure}[htbp]
    \centering
    \subfigure[]
    {\includegraphics[width=0.47\linewidth]{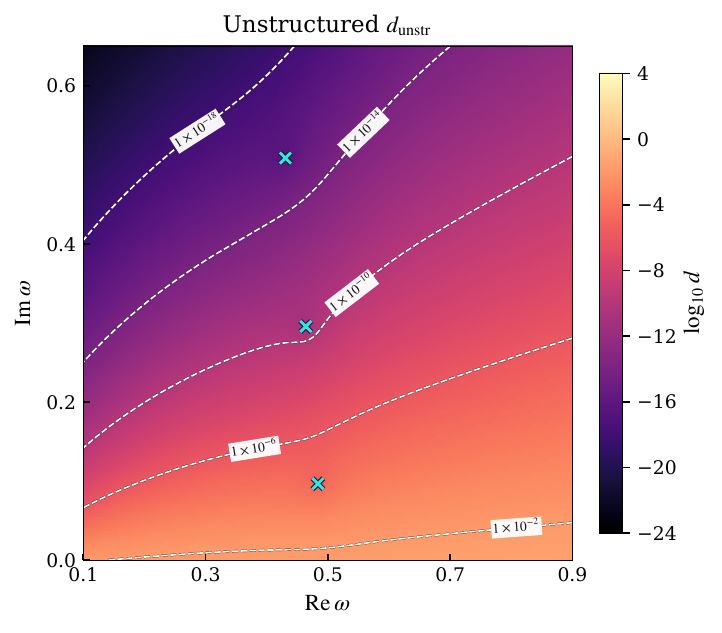}\label{unstr_d_RZ}}\hfill
    \subfigure[]
    {\includegraphics[width=0.47\linewidth]{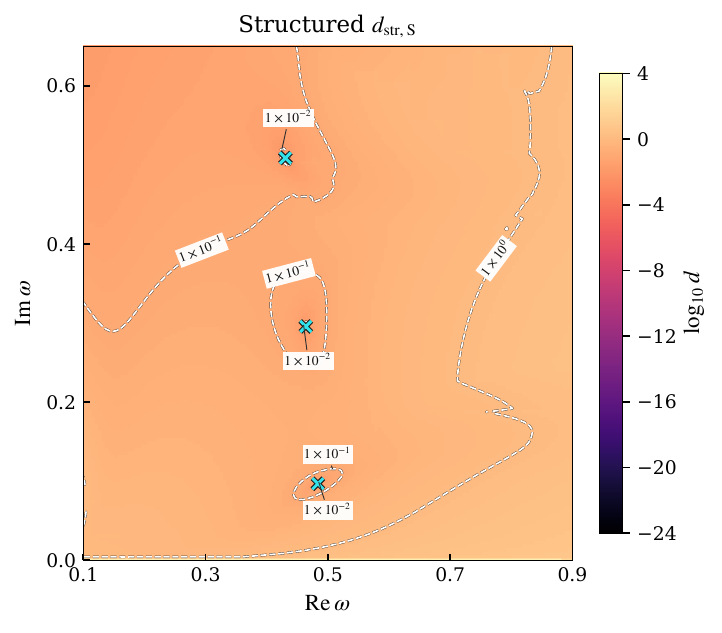}\label{str_d_S_RZ}}\\
    \subfigure[]
    {\includegraphics[width=0.47\linewidth]{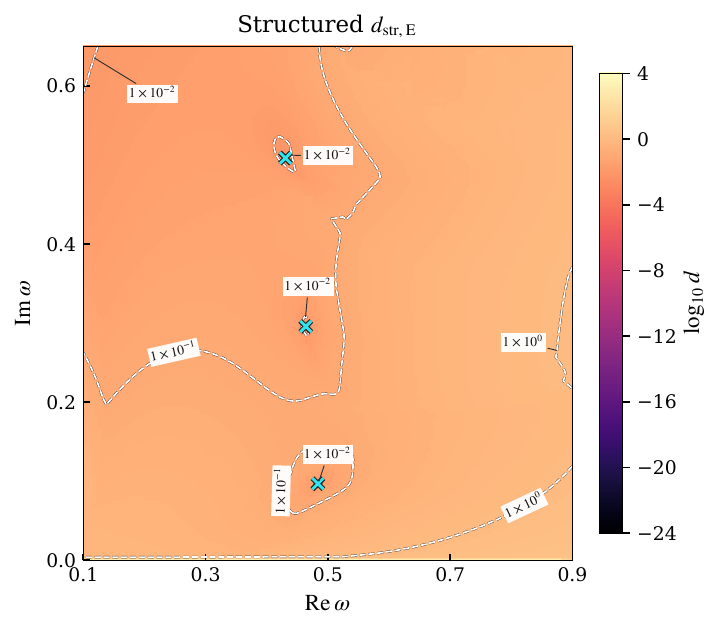}\label{str_d_E_RZ}}\hfill
    \subfigure[]
    {\includegraphics[width=0.47\linewidth]{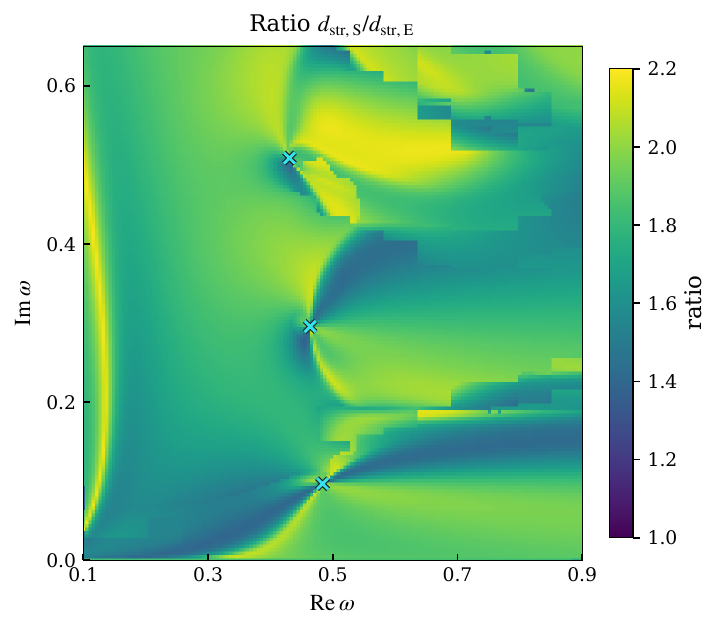}\label{Ratio_d_RZ}}
    \caption{Unstructured and RZ-structured pseudospectra of the Schwarzschild black hole for massless scalar perturbations $(s=0)$ with $m=4$. Panels (a)-(c) show the unstructured pseudospectrum $d_{\rm unstr}(\omega)$, the structured pseudospectrum $d_{\mathrm{str},\mathcal{S}}(\omega)$ defined with the HS norm, and $d_{\mathrm{str},\mathrm{E}}(\omega)$ that defined with the energy norm, respectively. These three panels use the same color legend for $\log_{10}d$, allowing a direct comparison of the spectral distances. The structured spectral distances are substantially larger than the unstructured one and vary more slowly over the complex-frequency plane, with noticeably smaller gradients. The dashed curves are selected $\varepsilon$-pseudospectrum contours. The crosses mark the original QNMs. Panel (d) shows the ratio $d_{\mathrm{str},\mathcal{S}}/d_{\mathrm{str},\mathrm{E}}$. The numerical set up is the same as Fig. \ref{fig:EFT_global_pseudospectrum}.}
    \label{fig:RZ_global_pseudospectrum}
\end{figure}

\begin{figure}[htbp]
    \centering
    \subfigure[]
    {\includegraphics[width=0.32\linewidth]{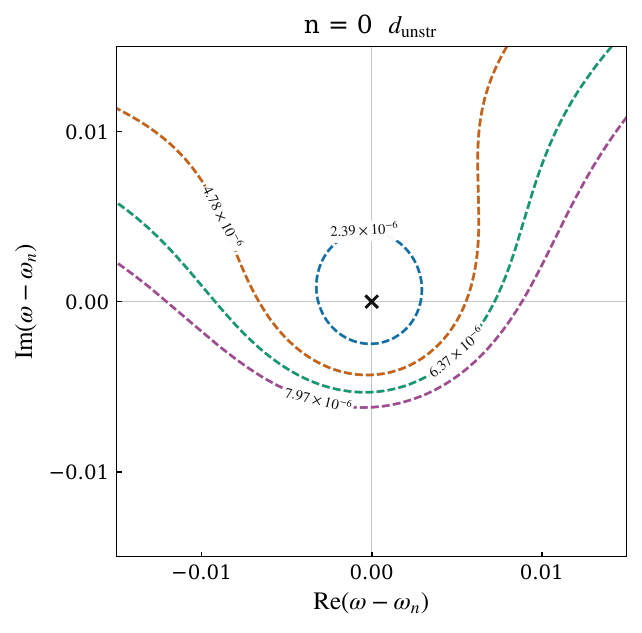}\label{d_unstr_EFT_n_0}}\hfill
    \subfigure[]
    {\includegraphics[width=0.32\linewidth]{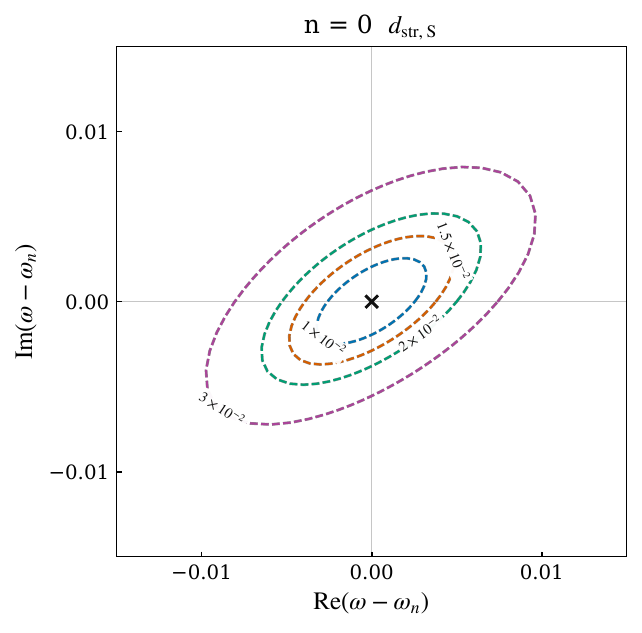}\label{d_str_EFT_S_n_0}}\hfill
    \subfigure[]
    {\includegraphics[width=0.32\linewidth]{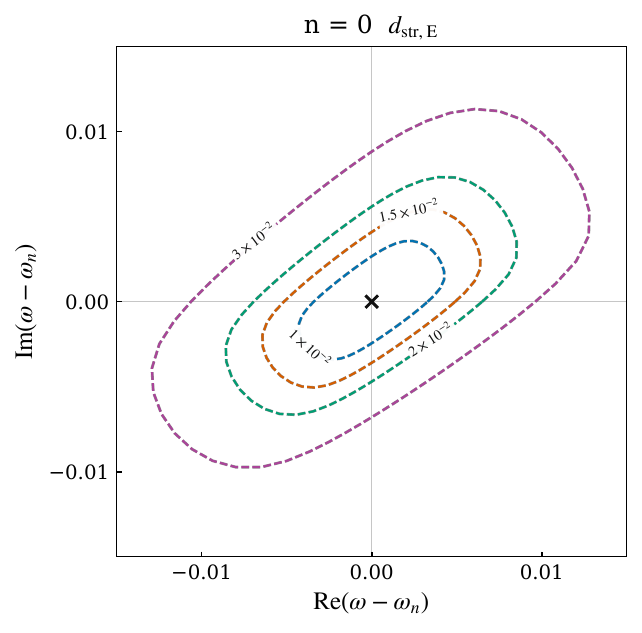}\label{d_str_EFT_E_n_0}}\\
    \subfigure[]
    {\includegraphics[width=0.32\linewidth]{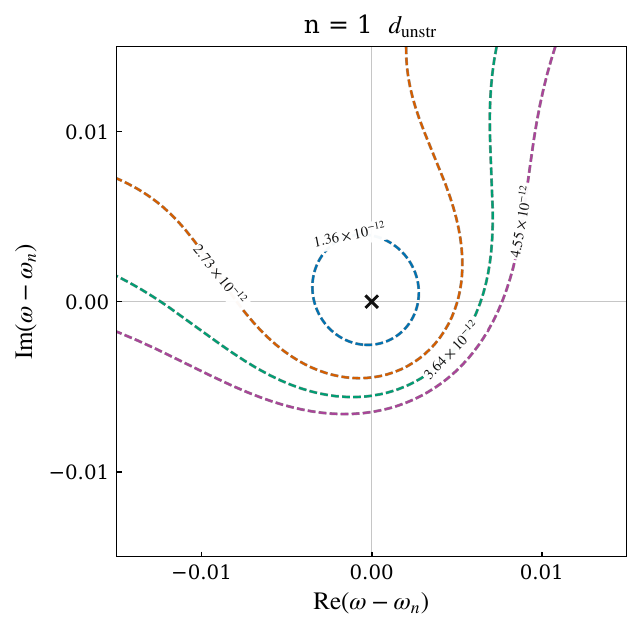}\label{d_unstr_EFT_n_1}}\hfill
    \subfigure[]
    {\includegraphics[width=0.32\linewidth]{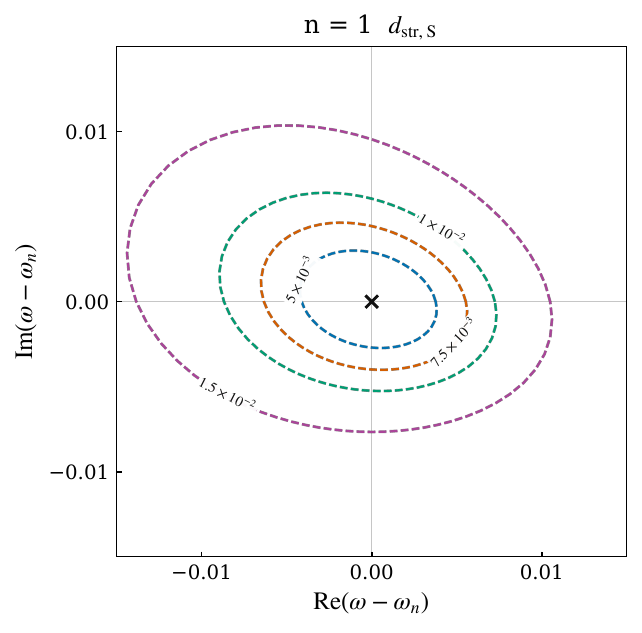}\label{d_str_EFT_S_n_1}}\hfill
    \subfigure[]
    {\includegraphics[width=0.32\linewidth]{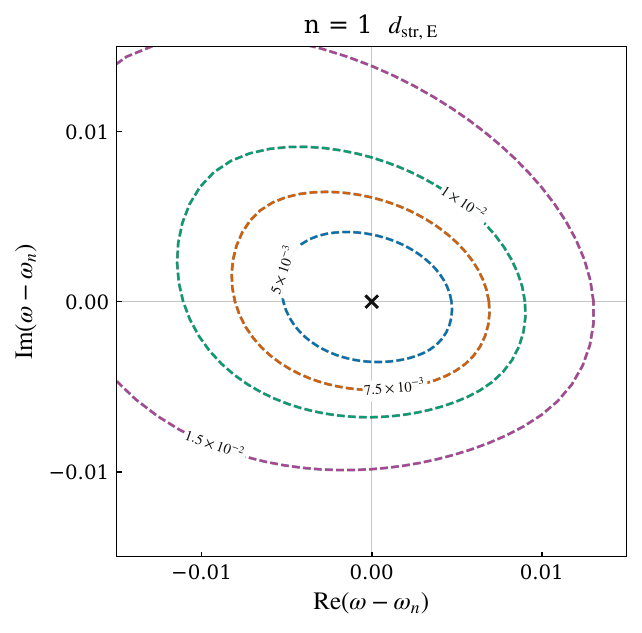}\label{d_str_EFT_E_n_1}}
    \caption{Unstructured and EFT-structured pseudospectra in the vicinity of the fundamental mode ($n=0$, upper panels) and the first overtone ($n=1$, lower panels). From left to right, the panels show the unstructured pseudospectrum, the structured pseudospectra $d_{\mathrm{str},\mathcal{S}}(\omega)$ and $d_{\mathrm{str},\mathrm{E}}(\omega)$, respectively. The spectrum is measured relative to the corresponding unperturbed one, so that the cross marking is located at the origin. The value of $\varepsilon$ is labeled on each contour with four contours being shown. The set up is the same as Fig. \ref{fig:EFT_global_pseudospectrum}.}
    \label{fig:EFT_local_pseudospectrum}
\end{figure}

\begin{figure}[htbp]
    \centering
    \subfigure[]
    {\includegraphics[width=0.32\linewidth]{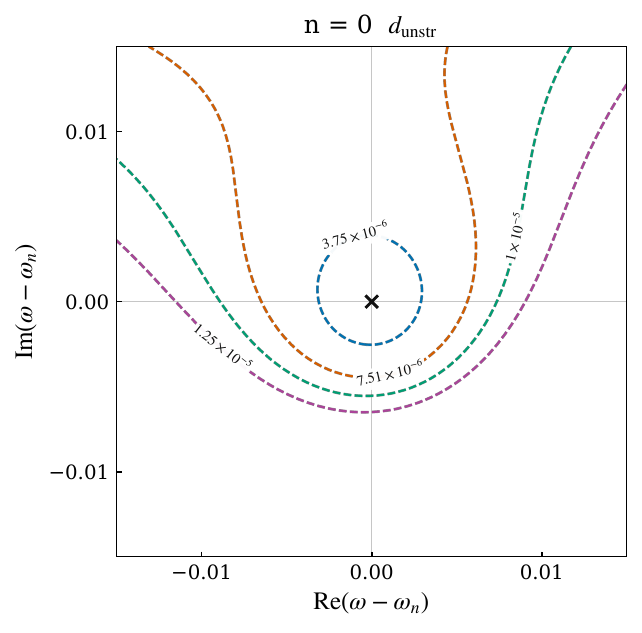}\label{d_unstr_RZ_n_0}}\hfill
    \subfigure[]
    {\includegraphics[width=0.32\linewidth]{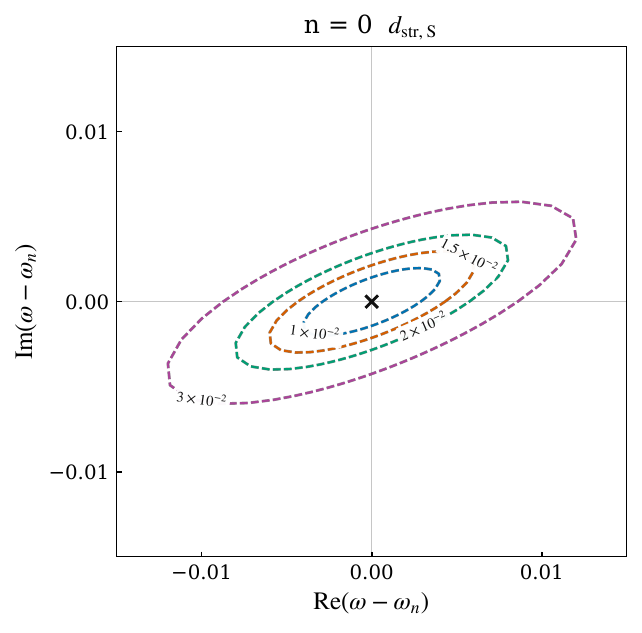}\label{d_str_RZ_S_n_0}}\hfill
    \subfigure[]
    {\includegraphics[width=0.32\linewidth]{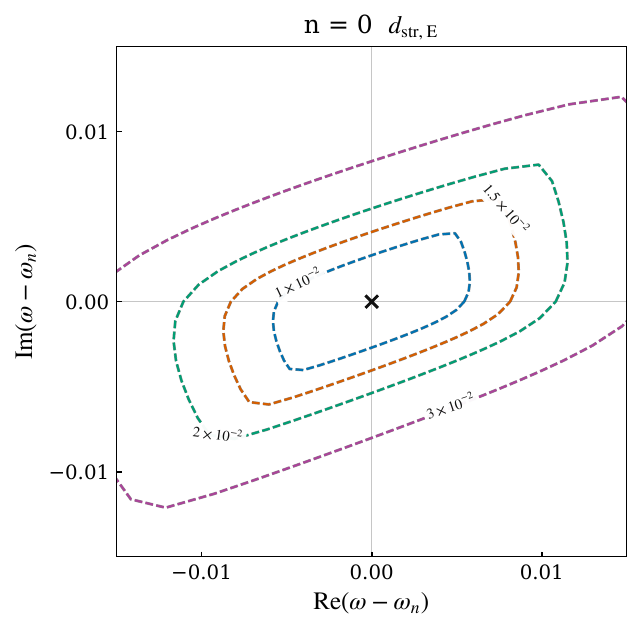}\label{d_str_RZ_E_n_0}}\\
    \subfigure[]
    {\includegraphics[width=0.32\linewidth]{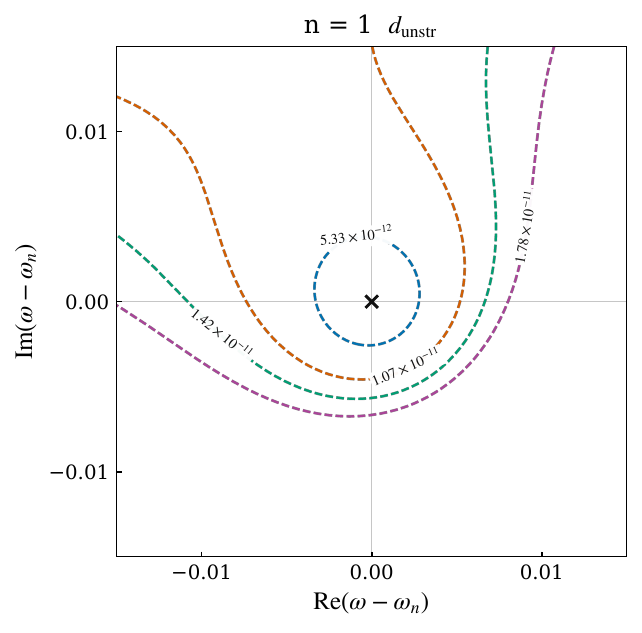}\label{d_unstr_RZ_n_1}}\hfill
    \subfigure[]
    {\includegraphics[width=0.32\linewidth]{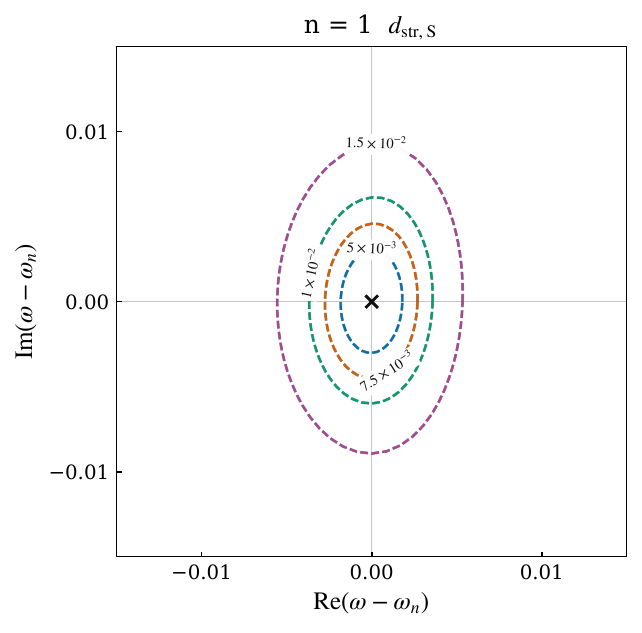}\label{d_str_RZ_S_n_1}}\hfill
    \subfigure[]
    {\includegraphics[width=0.32\linewidth]{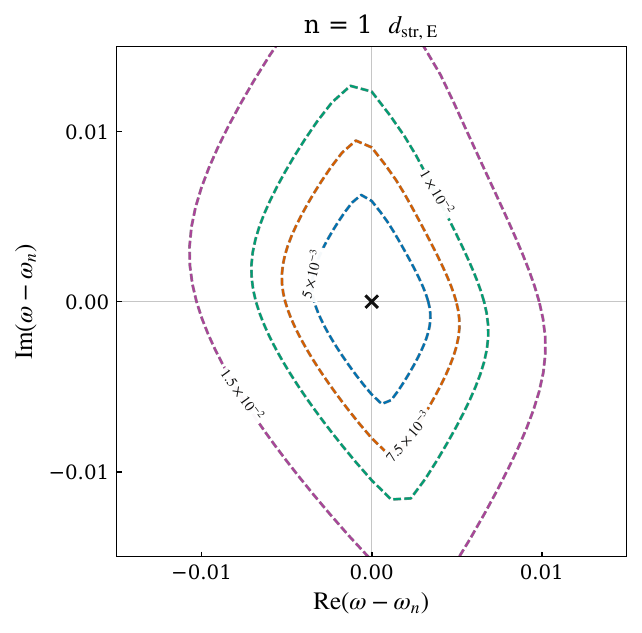}\label{d_str_RZ_E_n_1}}
    \caption{Unstructured and RZ-structured pseudospectra in the vicinity of the fundamental mode ($n=0$, upper panels) and the first overtone ($n=1$, lower panels). From left to right, the panels show the unstructured pseudospectrum, the structured pseudospectra $d_{\mathrm{str},\mathcal{S}}(\omega)$ and $d_{\mathrm{str},\mathrm{E}}(\omega)$, respectively. The spectrum is measured relative to the corresponding unperturbed one, so that the cross marking is located at the origin. The value of $\varepsilon$ is labeled on each contour with four contours being shown. The set up is the same as Fig. \ref{fig:RZ_global_pseudospectrum}.}
    \label{fig:RZ_local_pseudospectrum}
\end{figure}

We consider axial gravitational perturbations with $s=2$ for the EFT case and massless scalar perturbations with $s=0$ for the RZ case, and the dimension of perturbation space $\mathcal{S}$ is given by $m=4$. In Figs. \ref{fig:EFT_global_pseudospectrum} and \ref{fig:RZ_global_pseudospectrum}, the colored legend shows $\log_{10}d(\omega)$ for the corresponding spectral distance, while the dashed curves are representative contours $d(\omega)=\varepsilon$. The blue crosses denote the original Schwarzschild QNM spectra. Figures \ref{fig:EFT_local_pseudospectrum} and \ref{fig:RZ_local_pseudospectrum} further enlarge the neighborhoods of the pseudospectra for the fundamental mode $n=0$ and the first overtone $n=1$. In these figures, the modes have been shifted by the corresponding unperturbed QNMs, so that the mode under consideration is located at the origin. This representation allows us to compare more directly the magnitude and the directional dependence of the spectral stability under different classes of perturbations and overtones.

Figures \ref{unstr_d_EFT} and \ref{unstr_d_RZ} show the unstructured pseudospectra. It can be found that the unstructured spectral distance becomes extremely small over high imaginary part regions of the complex-frequency plane. Unstructured pseudospectra tell us that the instability of modes increases rapidly toward higher overtones. Given the spectrum migration distance, higher overtones require much less perturbation than the fundamental mode. However, the situation changes substantially once the perturbations are restricted to the prescribed physical spaces. Figures \ref{str_d_S_EFT}, \ref{str_d_E_EFT} and \ref{str_d_S_RZ}, \ref{str_d_E_RZ} show that the structured spectral distances are substantially larger than their unstructured counterparts over most of the displayed frequency regions. Correspondingly, at fixed $\varepsilon$, the structured pseudospectrum regions are much smaller than the corresponding unstructured pseudospectrum regions. Thus, although a given spectrum migration may be reached by an extremely small unrestricted operator perturbation, a much larger perturbation can be required once $\delta L$ is constrained to arise from the perturbation space $\mathcal{S}$.

This reduction clarifies the physical significance of the structured pseudospectrum. A large unstructured pseudospectrum establishes the existence of small perturbations in the full operator space capable of producing large spectrum migrations. Restricting $\delta L$ to a prescribed potential deformation space removes many of these perturbation directions. The strong suppression observed therefore shows that a substantial part of the unstructured spectrum instability is inaccessible within the physical perturbation sectors considered here.

The difference between the two structured spectral distances $d_{\mathrm{str},\mathcal{S}}(\omega)$ and $d_{\mathrm{str},\mathrm{E}}(\omega)$ is comparatively modest. For $N_{\text{c}}=120$, inequalities \eqref{inequality_between_three_d} gives
\begin{eqnarray}\label{theoretical_bound}
    1\leq\frac{d_{\mathrm{str},\mathcal{S}}}{d_{\mathrm{str},\mathrm{E}}}\leq\sqrt{N_{\text{c}}}\approx10.95\, .
\end{eqnarray}
As shown in Figs. \ref{Ratio_d_EFT} and \ref{Ratio_d_RZ}, the numerical ratio lies approximately between $1$ and $2.2$ in both two cases (considered frequency range), well within the above theoretical bound (\ref{theoretical_bound}). These two distances therefore introduce quantitative differences, including differences in the detailed contour shapes, while giving the same overall suppression relative to the unstructured pseudospectrum. In addition, this also ensures the reliability of the numerical results.

To demonstrate more refined results, we now examine the vicinities of the fundamental mode $n=0$ and the first overtone $n=1$. For the convenience of comparison, the observation window is the same. In the EFT case (see Fig. \ref{fig:EFT_local_pseudospectrum}), Fig. \ref{d_unstr_EFT_n_0} shows that the unstructured pseudospectrum contours change from closed to open at about order $10^{-6}$, whereas the structured pseudospectrum contours in Figs. \ref{d_str_EFT_S_n_0} and \ref{d_str_EFT_E_n_0} remain closed even at the $10^{-2}$ level. Correspondingly, spectrum displacements of order $|\omega-\omega_0|\sim10^{-2}$ have unstructured perturbation norm of order $10^{-6}$ but structured perturbation norm of order $10^{-2}$. The RZ case exhibits the same qualitative behavior (see Fig. \ref{fig:RZ_local_pseudospectrum}). Spectrum displacements of order $|\omega-\omega_0|\sim10^{-2}$ correspond to unstructured perturbation norm of order $10^{-5}$ and structured perturbation norm of order $10^{-2}$. Thus, in both two sectors, the fundamental mode is substantially more stable against the prescribed structured perturbations than its unstructured pseudospectrum would suggest. The contrast is even stronger for the first overtone. In the EFT case, the relevant unstructured pseudospectrum contours in Fig. \ref{d_unstr_EFT_n_1} occur at about $10^{-12}$ level, whereas the structured pseudospectrum contours in Figs. \ref{d_str_EFT_S_n_1}, \ref{d_str_EFT_E_n_1} remain at approximately $10^{-2}$ level. In the RZ case, the unstructured pseudospectrum contour in Fig. \ref{d_unstr_RZ_n_1} is at $10^{-12}\sim 10^{-11}$ level, while the structured pseudospectrum contours in Figs. \ref{d_str_RZ_S_n_1}, \ref{d_str_RZ_E_n_1} remain closed at perturbation scales of order $10^{-2}$. These results substantially modify the interpretation of overtone spectrum instability. The unstructured pseudospectrum shows an enormous increase in instability from the fundamental mode to the first overtone. However, the hierarchy between the fundamental mode and the first overtone is strongly reduced once the allowed perturbations are restricted. 

\begin{figure}[htbp]
    \centering
    \subfigure[]
    {\includegraphics[width=0.47\linewidth]{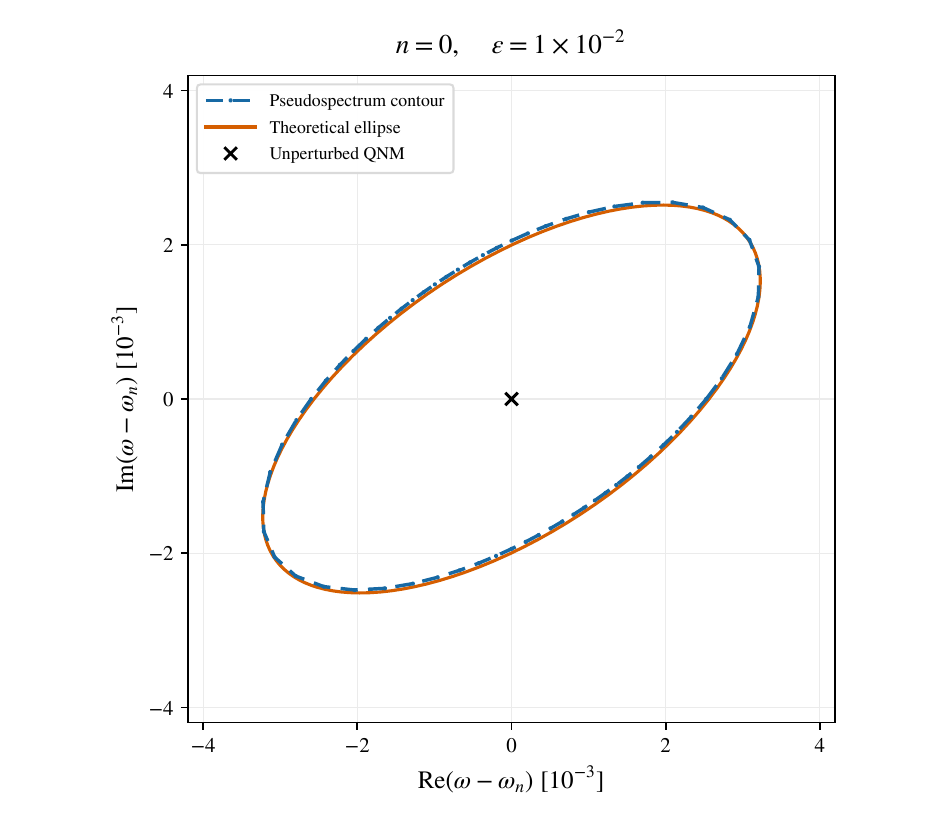}\label{EFT_n0_ellipse}}\hfill
    \subfigure[]
    {\includegraphics[width=0.47\linewidth]{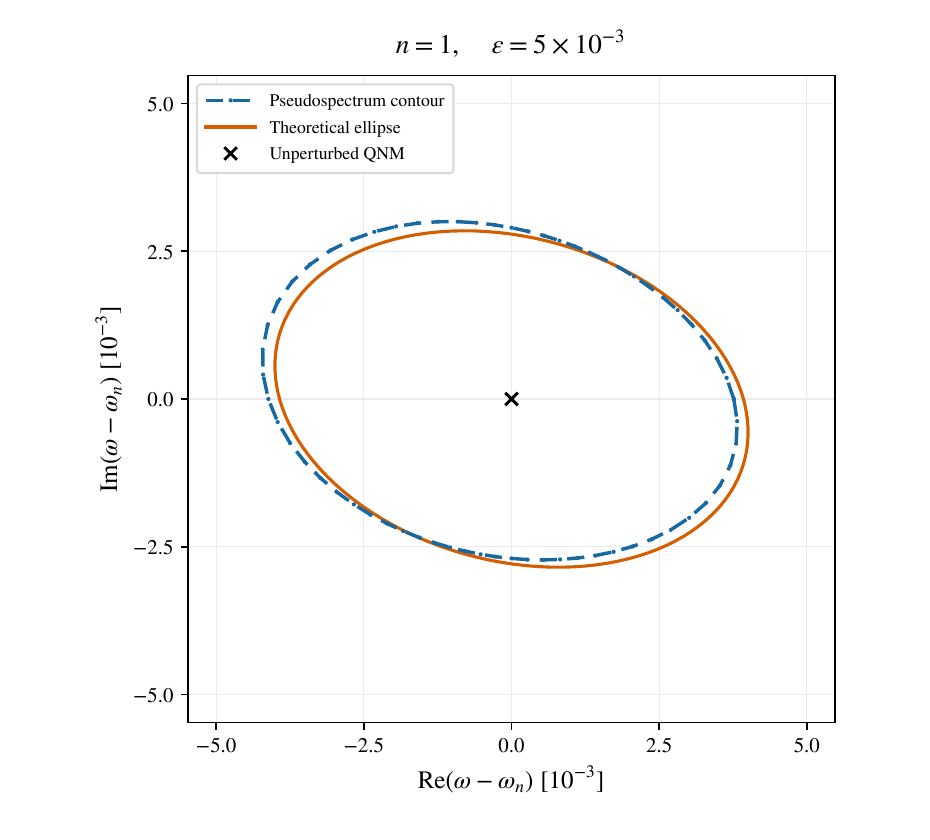}\label{EFT_n1_ellipse}}\\
    \subfigure[]
    {\includegraphics[width=0.47\linewidth]{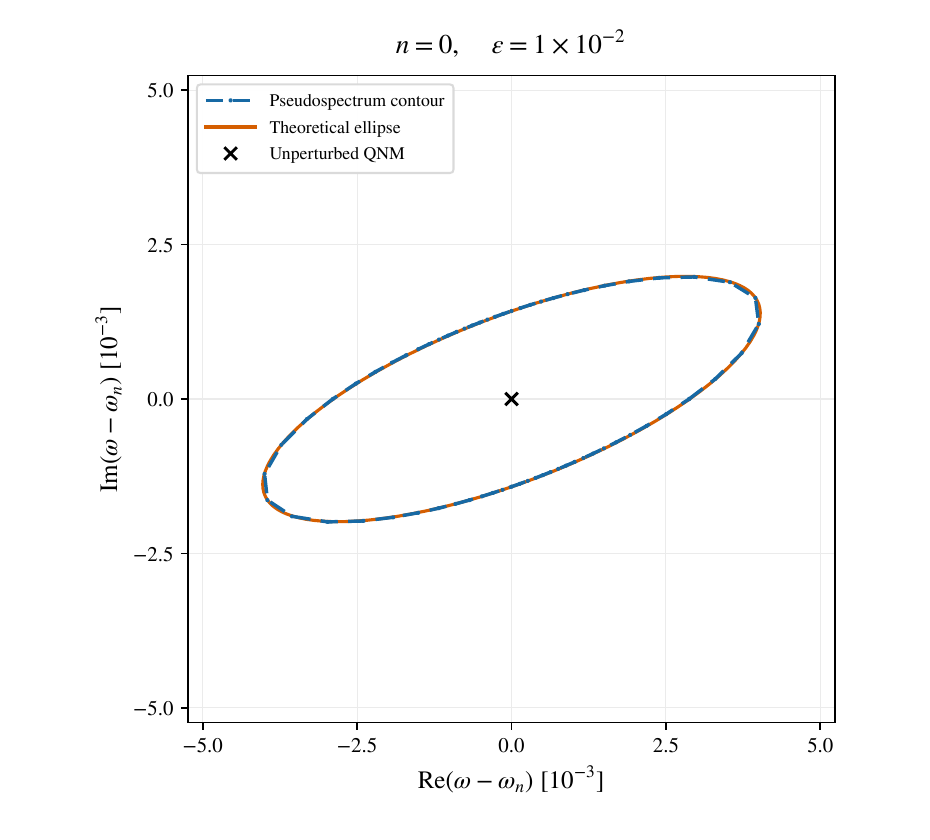}\label{RZ_n0_ellipse}}\hfill
    \subfigure[]
    {\includegraphics[width=0.47\linewidth]{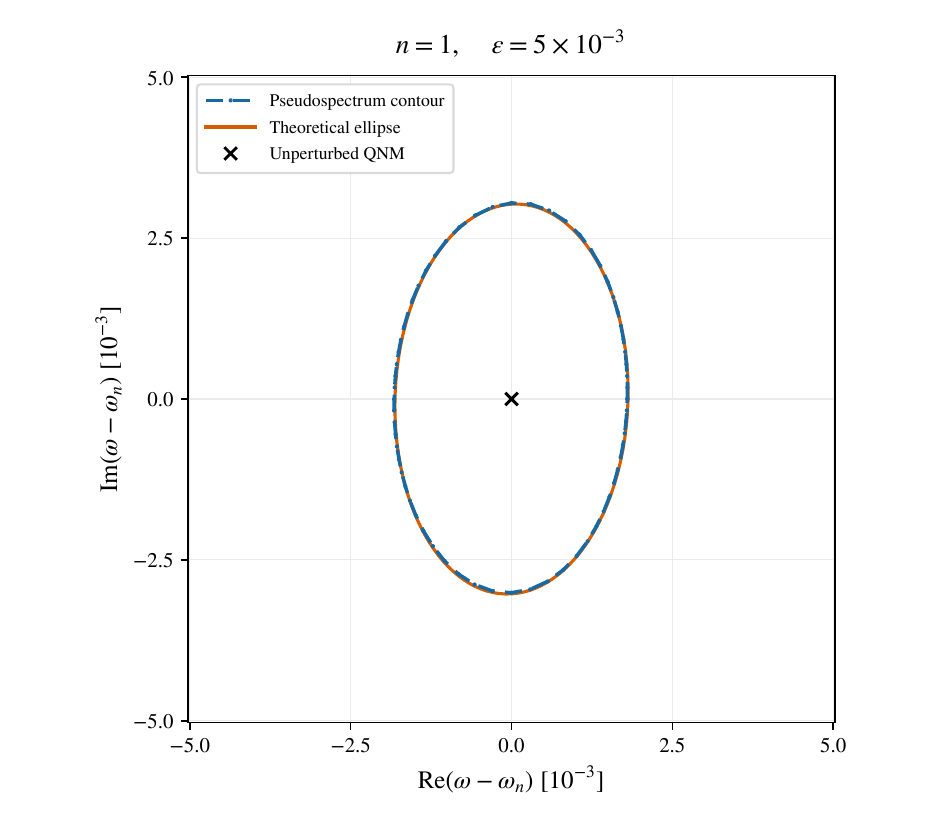}\label{RZ_n1_ellipse}}
    \caption{For the structural spectral distances                             $d_{\mathrm{str},\mathcal{S}}(\omega)$, the comparison between the ellipses in Eq. (\ref{eq:structured_ellipse}) and the pseudospectrum contours, where top two panels belong to EFT cases and bottom two panels belong to RZ cases. The dashed lines are the pseudospectrum contours while the solid lines are elliptical lines.}
    \label{fig:ellipse_and_pseudospectrum_contour}
\end{figure}

The local pseudospectrum contours also reveal a pronounced directional dependence of the structured response. Around an isolated simple QNM spectrum, the unstructured pseudospectrum contours become approximately circular for sufficiently small $\varepsilon$. The structured contours, by contrast, are anisotropic even in the small $\varepsilon$ regime. In the EFT case, the contours around the fundamental mode are elongated along an oblique direction, while those around the first overtone have a different orientation and aspect ratio. In the RZ case, the fundamental mode contours are likewise elongated obliquely, whereas the first overtone contours are elongated predominantly along the imaginary frequency direction. Therefore, the structured pseudospectrum encodes not only the magnitude of the accessible spectrum instability but also the preferred directions of QNM spectrum migration within a given perturbation space. Specifically, for the spectral distance $d_{\mathrm{str},\mathcal{S}}(\omega)$, the local pseudospectrum contours asymptotically approach ellipses as the perturbation scale tends to zero. This is summarized as the theorem in Appendix \ref{app: elliptical_proof}. For $d_{\mathrm{str},\mathcal{S}}(\omega)$, in Fig. \ref{fig:ellipse_and_pseudospectrum_contour}, we show a comparison between the ellipses in Eq. (\ref{eq:structured_ellipse}) and the actual pseudospectrum contours, where the corresponding $\varepsilon$ is explicitly indicated in each panels. We can see that when $\varepsilon$ is small, the pseudospectrum contours indeed approach the ellipses. This further demonstrates the reliability of our numerical results.

\section{Conclusions and discussion}\label{sec: conclusions}
In this work, we investigate the structured pseudospectra of Schwarzschild black holes by restricting the effective potential deformations to physically motivated perturbation spaces. In particular, we consider two classes of perturbations inspired by effective field theory (EFT) and the Rezzolla--Zhidenko (RZ) parametrization, respectively. To characterize the pseudospectra, we have employed two different norms and established a hierarchy relating the corresponding structured spectral distances to the conventional unstructured one [refer to inequalities (\ref{inequality_between_three_d})].

Our numerical results show that the spectrum instability predicted by the conventional pseudospectrum is substantially suppressed when the perturbations are restricted to the considered physical spaces. This behavior is observed for both the fundamental mode and the first overtone, with a particularly pronounced contrast for the latter (see Fig. \ref{fig:EFT_local_pseudospectrum} and Fig. \ref{fig:RZ_local_pseudospectrum}). Although the unstructured pseudospectrum suggests that extremely small operator perturbations can produce appreciable spectral displacements, considerably larger perturbations are required within the EFT- and RZ-motivated sectors. These findings indicate that much of the strong spectrum instability revealed by unrestricted perturbations may not be accessible within a given class of physical deformations.

Another important finding concerns the geometry of the pseudospectrum contours. For sufficiently small $\varepsilon$, the unstructured pseudospectrum approaches a circular shape around an isolated simple QNM. In contrast, the structured pseudospectra retain approximately elliptical-like contours, whose orientations and elongations depend on the particular QNM and the underlying perturbation model. For the structured spectral distance $d_{\mathrm{str},\mathcal{S}}(\omega)$, the contour is exactly the asymptotic ellipse. Mathematical proof for this  can be found in Appendix \ref{app: elliptical_proof}, while the numerical verification can be found in Fig. \ref{fig:ellipse_and_pseudospectrum_contour}. The difference between the unstructured pseudospectrum and the structured pseudospectrum reveals that physically constrained perturbations introduce preferred directions of spectral migration, even in the small perturbation regime. Therefore, the structured pseudospectrum provides information not only about the magnitude of spectrum instability but also about its directional dependence, which is absent from the locally circular structure of the unstructured pseudospectrum.

There are some things that can be further explored based on the present work. Here, we list some investigations that can be done in the future. Unstructured (ordinary) pseudospectrum of Kerr black hole has been obtained in~\cite{Cai:2025irl,Zhang:2026cjq}, which motivate us to study the structured pseudospectrum of Kerr black hole. Total transmission modes (TTMs) which are another important modes on black hole perturbation theory can also be considered~\cite{Yu:2026rku,Dong:2026zxy,Zhou:2025xdo,Wu:2026hvf,Qian:2025occ,Tuncer:2025dnp,Rosato:2025byu,Chen:2026ane}. The structured pseudospectrum for TTMs is a very natural next step for studying.

\appendix

\section*{Acknowledgement}
This work is supported by the National Natural Science Foundation of China with grants No. 12505067, No. 12475063.

\appendix

\section{Physically motivated construction of the structured perturbation space}\label{app: construction_S}
In this appendix, we construct the structured perturbation space $\mathcal{S}$ from physically motivated corrections to the effective potential of Schwarzschild black hole. Our main requirement is that the background geometry (Schwarzschild black hole), the horizon position, and the Schwarzschild tortoise coordinate remain unchanged (Or it can be valid after some approximations.), so that modification only enters into the effective potential.

One suitable example is provided by the  effective field theory (EFT) containing the higher-curvature operator $\widetilde{\mathcal{C}}$ and $\mathcal{C}$, where $\widetilde{\mathcal{C}}=R_{\alpha\beta\gamma\delta}\widetilde{R}^{\alpha\beta\gamma\delta}$, and $\mathcal{C}=R_{\alpha\beta\gamma\delta}R^{\alpha\beta\gamma\delta}$~\cite{Cardoso:2018ptl}. The modified axial gravitational perturbation equation, obtained from its GR counterpart through the modification in Eq. (A6) of Ref. \cite{Cardoso:2018ptl}, can be written as
\begin{eqnarray}
    \frac{\mathrm{d}^2\Psi_{-}}{\mathrm{d}x^2}+\Big[\omega^2-f(r)\big(V_{-}^{\rm GR}+\epsilon_2 V_{-}^{\epsilon_2}\big)\Big]\Psi_{-}=0\, , \quad \text{with} \quad \frac{\mathrm{d} x}{\mathrm{d} r}=\frac{1}{f(r)}\, ,
\end{eqnarray}
in which $V_{-}^{\mathrm{GR}}$ is Regge--Wheeler potential (axial gravitational perturbations $s=2$) and the modified term satisfies $V_{-}^{\epsilon_2}\propto 1/r^{10}$. Therefore, the EFT correction modifies only the Schwarzschild effective potential while preserving the original tortoise coordinate. Introducing the dimensionless radial variable $\sigma=r_{\text{h}}/r$ [see Eq. (\ref{compact_hyperboloidal_coordinates})], the above modification $f(r)V_{-}^{\epsilon_2}$ can be proportional to the following function
\begin{eqnarray}
    \phi_{10}(\sigma)=(1-\sigma)\sigma^{10}\, .
\end{eqnarray} 
Accordingly, inspired by such a construction, the EFT-matched perturbation space is defined as
\begin{eqnarray}\label{S_EFT}
    \mathcal{S}_{\mathrm{EFT}}^{(\mathcal{N})}:=\operatorname{span}\Big\{\phi_{10},\phi_{11},\cdots,\phi_{j},\cdots,\phi_\mathcal{N}\Big\}\, ,\quad \text{with} \quad \phi_{j}=(1-\sigma)\sigma^{j}\, .
\end{eqnarray}
It is not difficult to find that the dimension of $\mathcal{S}_{\mathrm{EFT}}^{(\mathcal{N})}$ is $m=\mathcal{N}-9$ with $\mathcal{N}\geq10$. The higher-order basis functions with $j>10$ are introduced as EFT-inspired extensions of the leading $j=10$ correction. Physically, they may represent subleading strong-field effects associated with higher-dimensional curvature operators or operators containing additional derivatives. Such contributions are expected to decay more rapidly away from the black hole. We emphasize that, unlike the $j=10$ mode, these higher-$j$ components are not assumed to have a one-to-one correspondence with specific EFT operators, but rather provide a controlled parametrization of possible EFT corrections. A perturbation space similar to (\ref{S_EFT}) appears in~\cite{Cardoso:2019mqo}.

Another reliable construction for the structured perturbation space originate from Rezzolla--Zhidenko (RZ) parametrization~\cite{Rezzolla:2014mua}. Consider a static and spherically symmetric spacetime written as~\cite{Rezzolla:2014mua}
\begin{eqnarray}\label{RZ_metric}
    \mathrm{d}s^2=-N^2(r)\mathrm{d}t^2+\frac{B^2(r)}{N^2(r)}\mathrm{d}r^2+r^2 \mathrm{d}\Omega^2 \, .
\end{eqnarray}
The Rezzolla--Zhidenko (RZ) parametrization introduces the compact radial coordinate $y=1-r_{\text{h}}/r$, where $r_{\text{h}}$ denotes the radius of the event horizon, such that $y=0$ corresponds to the horizon and $y=1$ to spatial infinity. The metric function $N^2$ in metric (\ref{RZ_metric}) is written as
\begin{eqnarray}
    N^2(y)=y A(y)\, ,
\end{eqnarray}
where the function $A(y)$ is
\begin{eqnarray}
    A(y)=1-\epsilon(1-y)+(a_0-\epsilon)(1-y)^2+\widetilde A(y)(1-y)^3 \, .
\end{eqnarray}
The function $B(y)$ in metric (\ref{RZ_metric}) is
\begin{eqnarray}
    B(y)=1+b_0(1-y)+\widetilde B(y)(1-y)^2 \, .
\end{eqnarray}
The functions $\widetilde A(y)$ and $\widetilde B(y)$ describing the strong-field region are represented by continued fractions, namely
\begin{eqnarray}\label{tilde_A_and_tilde_B}
    \widetilde{A}(y)=\frac{a_1}{1+\dfrac{a_2y}{1+\dfrac{a_3y}{1+\cdots}}}\, ,\quad \text{and} \quad \widetilde{B}(y)=\frac{b_1}{1+\dfrac{b_2y}{1+\dfrac{b_3y}{1+\cdots}}}\, .
\end{eqnarray}
The parameter $\epsilon$ is related to the mass $M$ and the horizon radius $r_\text{h}$ through
\begin{eqnarray}
    \epsilon=\frac{2M-r_\text{h}}{r_\text{h}}\, .
\end{eqnarray}
where the Schwarzschild black hole corresponds to $\epsilon=a_0=a_1=\cdots=b_0=b_1=\cdots=0$.

We choose a special subspace in the RZ parameter space that satisfies conditions that $\epsilon=0$ and $B=A$. These two conditions ensure that the horizon and tortoise coordinates remain unchanged. Furthermore, we consider the massless scalar field $(s=0)$ which satisfies $\square\Phi=0$. Using the RZ metric (\ref{RZ_metric}), one obtains the difference between the effective potential of the Schwarzschild metric and the RZ metric, which is given by
\begin{eqnarray}
    \delta V_{\mathrm{RZ}}(\sigma)=\frac{\ell(\ell+1)}{4M^2}\Big[a_0(1-\sigma)\sigma^4+(1-\sigma)\sigma^5\widetilde A(\sigma)\Big]\, .
\end{eqnarray}
where we have used the relation between $y$ and $\sigma$, i.e., $y=1-\sigma$. Here, $\widetilde{A}(\sigma)\equiv\widetilde{A}(y(\sigma))$. Generally, if $\widetilde A(\sigma)$ is analytic in a complex neighborhood of $[0,1]$, it admits a uniformly and exponentially convergent expansion in shifted Chebyshev polynomials (Bernstein theorem). Therefore, we approximate the RZ deformation as
\begin{eqnarray}
    \widetilde A(\sigma)\simeq\sum_{k=0}^{N} c_k T_k(2\sigma-1)\, ,
\end{eqnarray}
where $T_k(2\sigma-1)$ is the shifted Chebyshev polynomials. Accordingly, inspired by the construction, the RZ-matched perturbation space is defined as
\begin{eqnarray}\label{S_RZ}
    \mathcal{S}^{(\mathcal{N})}_{\rm RZ}:=\operatorname{span}\Big\{(1-\sigma)\sigma^4,\big\{(1-\sigma)\sigma^5T_k(2\sigma-1)\big\}_{k=0}^{\mathcal{N}}\Big\}\, ,
\end{eqnarray}
where the dimension of $\mathcal{S}^{(\mathcal{N})}_{\rm RZ}$ is $\mathcal{N}+2$.

\section{Method for getting structured distances}\label{app: structured_distance}
In this appendix, we propose the method of getting structured distances $d_{\mathrm{str},\mathcal{S}}(\omega)$ and $d_{\mathrm{str},\mathrm{E}}(\omega)$ for given $\omega\in\mathbb{C}$. Note that in this appendix, we always use the matrix version of the operator constructed using Chebyshev--Lobatto grid. Given a resolution $N_\text{c}$, the dimension of the matrix $\mathbf{L}$ is $n_{\text{U}}=2(N_{\text{c}}+1)$.

First, we concentrate on $d_{\mathrm{str},\mathcal{S}}(\omega)$. Since we have Eq. (\ref{delta_V_norm}), the structured spectral distance is equivalent to
\begin{eqnarray}\label{structured_spectral_distance_2_S}
    d_{\mathrm{str},\mathcal{S}}(\omega)=\min_{\mathbf{c}\in\mathbb{R}^m}\sqrt{\mathbf{c}^{\mathrm{T}}\mathbf{G}\mathbf{c}} \quad \text{subject to} \quad \sigma_{\min}\Big(\omega \mathbf{I}-\mathbf{L}+\frac{1}{\mathrm{i}}\sum_{k=1}^{m}c_k\mathbf{M}_k\Big)=0\, ,
\end{eqnarray}
where Eq. (\ref{delta_L_in_terms_of_phi}) is used. Since the basis functions $\{\phi_k\}_{k=1}^{m}$ of $\mathcal{S}$ are not necessarily orthogonal, the perturbation norm is characterized by the Gram matrix $\mathbf{G}$ [see Eq. (\ref{Gram_definition})] which has been induced from the original energy inner product (\ref{eq:appB_energy_inner}). It is convenient to transform the coefficient vector into an orthonormal representation. We perform the Cholesky decomposition of the Gram matrix $\mathbf{G}$ as follow
\begin{eqnarray}\label{Cholesky_decomposition_G}
    \mathbf{G}=\mathbf{R}^{\mathrm T}\mathbf{R}\, ,
\end{eqnarray}
where $\mathbf{R}$ is an upper triangular matrix with dimension $m\times m$, and is invertible. Introducing a new variable $\mathbf{y}=\mathbf{R}\mathbf{c}$, the norm of the perturbation $\delta V$ in $\mathcal{S}$ becomes
\begin{eqnarray}
    \|\delta V\|_{\mathcal S}=\sqrt{\mathbf c^{\mathrm{T}}\mathbf{G}\mathbf c}=\sqrt{\mathbf c^{\mathrm{T}}\mathbf{R}^{\mathrm{T}}\mathbf{R}\mathbf c}=\|\mathbf y\|_{2}\, .
\end{eqnarray}
Therefore, the structured perturbation problem can be reformulated in terms of the Euclidean norm of $\mathbf y$. Since $\mathbf c=\mathbf{R}^{-1}\mathbf y$, the perturbed matrix [see Eq. (\ref{delta_L_in_terms_of_phi})] can be rewritten as
\begin{eqnarray}
    \delta \mathbf{L}=-\frac{1}{\mathrm i}\sum_{k=1}^{m}c_k\mathbf{M}_k=-\frac{1}{\mathrm i}\sum_{j=1}^{m}y_j\widetilde{\mathbf{M}}_j\, ,
\end{eqnarray}
where the transformed perturbation matrices are defined as
\begin{eqnarray}\label{transformed_perturbation_matrices}
    \widetilde{\mathbf{M}}_{j}=\sum_{k=1}^{m}(\mathbf{R}^{-1})_{kj}\mathbf{M}_k \, .
\end{eqnarray}
Consequently, the structured spectral distance (\ref{structured_spectral_distance_2_S}) can be equivalently expressed as
\begin{eqnarray}\label{dstr_subject_to_sigma_min}
    d_{\mathrm{str},\mathcal{S}}(\omega)=\min_{\mathbf y\in\mathbb R^m}\|\mathbf y\|_{2}\, ,\quad \text{subject to}\quad \sigma_{\min}\Big(\omega \mathbf{I}-\mathbf{L}+\frac{1}{\mathrm{i}}\sum_{j=1}^{m}y_j\widetilde{\mathbf{M}}_{j}\Big)=0\, .
\end{eqnarray}
This transformation converts the weighted perturbation norm into the standard Euclidean norm, which is more suitable for numerical optimization. The core of Eq. (\ref{dstr_subject_to_sigma_min}) is to find a minimum real perturbation $\mathbf y$ constrained by a given set function subspace for a fixed complex frequency $\omega$, such that $\omega$ happens to be the characteristic frequency of the perturbed matrix.

For the convenience of marking, given $\omega\in\mathbb{C}$, we define 
\begin{eqnarray}\label{A_0_and_B_j}
    \mathbf{A}_0(\omega):=\omega \mathbf{I}-\mathbf{L}\, ,\quad \text{and}\quad  \mathbf{B}_j:=-\mathrm{i}\widetilde{\mathbf{M}}_{j}\, ,
\end{eqnarray}
then the structured spectral distance (\ref{dstr_subject_to_sigma_min}) can be written as
\begin{eqnarray}\label{dstr_subject_to_sigma_min_2}
    d_{\mathrm{str},\mathcal{S}}(\omega)=\min_{\mathbf y\in\mathbb R^m}\|\mathbf y\|_{2}\, ,\quad \text{subject to}\quad \sigma_{\min}\Big[\mathbf{A}(\mathbf{y})\Big]=0\, ,\quad \text{with} \quad \mathbf{A}(\mathbf{y})=\mathbf{A}_0+\sum_{j=1}^my_j\mathbf{B}_j\, ,\quad y_j\in\mathbb{R}\, .
\end{eqnarray}
We have noted $\sigma_{\min}\Big[\mathbf{A}(\mathbf{y})\Big]=0$ is equivalent to there exits no-zero right null vector $\mathbf{v}$ such that $\mathbf{A}(\mathbf{y})\mathbf{v}=0$, in which $\mathbf{v}^{\dagger}\mathbf{v}=1$. Accordingly, the optimization problem (\ref{dstr_subject_to_sigma_min_2}) is equivalent to
\begin{eqnarray}\label{dstr_subject_to_sigma_min_3}
    d_{\mathrm{str},\mathcal{S}}(\omega)=\min_{\mathbf{y},\mathbf{v}}\sqrt{\mathbf y^{\mathrm{T}}\mathbf{y}}\, ,\quad \text{subject to}\quad \mathbf{A}(\mathbf{y})\mathbf{v}=0\, ,\quad \mathbf{v}^{\dagger}\mathbf{v}=1\, ,\quad \mathbf{y}\in\mathbb{R}^m\, ,\quad \mathbf{v}\in\mathbb{C}^{n_{\text{U}}}\, .
\end{eqnarray}
Because $\mathbf{A}(\mathbf{y})$ and $\mathbf{v}$ are still complex, we should turn the above problem into a pure real problem. One writes $\mathbf{v}=\mathbf{a}+\mathrm{i}\mathbf{b}$, and $\mathbf{A}(\mathbf{y})=\mathbf{A}_{\text{R}}(\mathbf{y})+\mathrm{i}\mathbf{A}_{\text{I}}(\mathbf{y})$. From the condition $\mathbf{A}(\mathbf{y})\mathbf{v}=0$, we obtain two real conditions given by
\begin{eqnarray}
    \mathbf{A}_{\text{R}}(\mathbf{y})\mathbf{a}-\mathbf{A}_{\text{I}}(\mathbf{y})\mathbf{b}=0\, ,\quad \text{and} \quad \mathbf{A}_{\text{R}}(\mathbf{y})\mathbf{b}+\mathbf{A}_{\text{I}}(\mathbf{y})\mathbf{a}=0\, ,\quad \text{with}\quad \mathbf{a}\in\mathbb{R}^{n_{\text{U}}}\, ,\quad \text{and} \quad \mathbf{b}\in\mathbb{R}^{n_{\text{U}}}\, .
\end{eqnarray}
Furthermore, we need to remove the phase degree of freedom of $\mathbf{v}$. For example, we select the component with the highest absolute value in the range of $\mathbf{v}$, and specify $\operatorname{Im}(v_k)=0$ which is $b_k=0$. In summary, the complete optimization problem for $d_{\mathrm{str},\mathcal{S}}(\omega)$ is given as follow
\begin{eqnarray}\label{complete_optimization_problem_dstr_S}
    d_{\mathrm{str},\mathcal{S}}(\omega)&=&\min_{\mathbf{y}\in\mathbb{R}^m,\mathbf{a},\mathbf{b}\in\mathbb{R}^{n_{\text{U}}}}\sqrt{\mathbf y^{\mathrm{T}}\mathbf{y}}\, ,\quad \text{subject to}\nonumber\\
    &&\mathbf{A}_{\text{R}}(\mathbf{y})\mathbf{a}-\mathbf{A}_{\text{I}}(\mathbf{y})\mathbf{b}=0\, ,\quad \mathbf{A}_{\text{R}}(\mathbf{y})\mathbf{b}+\mathbf{A}_{\text{I}}(\mathbf{y})\mathbf{a}=0\, ,\quad \mathbf{a}^{\mathrm{T}}\mathbf{a}+\mathbf{b}^{\mathrm{T}}\mathbf{b}=1\, ,\quad\text{and}\quad b_k=0\, .
\end{eqnarray}

Now, we present the numerical method for evaluating the energy norm structured spectral distance $d_{\mathrm{str},\mathrm{E}}(\omega)$. The definition of this distance is given by Eq. (\ref{structured_spectral_distance_E}). The norm of the structured operator perturbation becomes
\begin{eqnarray}\label{delta_L_norm_E}
    \|\delta \mathbf{L}\|_{\mathrm{E}}=\| \widetilde{\delta\mathbf{L}}\|_{2}=\| \mathbf{W}(\delta \mathbf{L})\mathbf{W}^{-1}\|_2 \, .
\end{eqnarray}
For convenience, we define the transformed perturbation matrices
\begin{eqnarray}
    \mathbf{C}_k=-\mathrm{i}\mathbf{W}\mathbf{M}_k\mathbf{W}^{-1}\, .
\end{eqnarray}
Using Eq. (\ref{delta_L_in_terms_of_phi}) and Eq. (\ref{delta_L_norm_E}), the above equation gives the energy norm of the structured perturbation
\begin{eqnarray}\label{energy_norm_structured_perturbation}
    \|\delta \mathbf{L}\|_{\mathrm{E}}=\Big\|\sum_{k=1}^{m}c_k\mathbf{C}_k\Big\|_2 \, .
\end{eqnarray}
Similar to the processes from Eq. (\ref{A_0_and_B_j}) to Eq. (\ref{complete_optimization_problem_dstr_S}), we ultimately obtain the complete optimization problem for $d_{\mathrm{str},\mathrm{E}}(\omega)$ as follow
\begin{eqnarray}\label{complete_optimization_problem_dstr_E}
    d_{\mathrm{str},\mathrm{E}}(\omega)&=&\min_{\mathbf{c}\in\mathbb{R}^m,\mathbf{a},\mathbf{b}\in\mathbb{R}^{n_{\text{U}}}}\Big\|\sum_{k=1}^{m}c_k\mathbf{C}_k\Big\|_2\, ,\quad \text{subject to}\nonumber\\
    &&\mathbf{D}_{\text{R}}(\mathbf{c})\mathbf{a}-\mathbf{D}_{\text{I}}(\mathbf{c})\mathbf{b}=0\, ,\quad \mathbf{D}_{\text{R}}(\mathbf{c})\mathbf{b}+\mathbf{D}_{\text{I}}(\mathbf{c})\mathbf{a}=0\, ,\quad \mathbf{a}^{\mathrm{T}}\mathbf{a}+\mathbf{b}^{\mathrm{T}}\mathbf{b}=1\, ,\quad\text{and}\quad b_k=0\, .
\end{eqnarray}
For the above optimization problem, the vector $\mathbf{a}+\mathrm{i}\mathbf{b}$ is no-zero right null vector of the matrix $\mathbf{D}$ which is given by
\begin{eqnarray}\label{D_matrix}
    \mathbf{D}(\mathbf{c})=\mathbf{A}_0-\delta \mathbf{L}=\mathbf{A}_0-\mathrm{i}\sum_{k=1}^mc_k\mathbf{M}_k\, ,
\end{eqnarray}
where $\mathbf{A}_0$ is from Eqs. (\ref{A_0_and_B_j}). In addition, $\mathbf{D}(\mathbf{c})=\mathbf{D}_{\text{R}}(\mathbf{c})+\mathrm{i}\mathbf{D}_{\text{I}}(\mathbf{c})$, namely $\mathbf{D}_{\text{R}}$ and $\mathbf{D}_{\text{I}}$ are the real and imaginary parts of $\mathbf{D}$, respectively.

By solving the two optimization problems (\ref{complete_optimization_problem_dstr_S}) and (\ref{complete_optimization_problem_dstr_E}), we obtain the structured spectral distances $d_{\mathrm{str},\mathcal{S}}(\omega)$ and $d_{\mathrm{str,E}}(\omega)$, respectively, and thereby derive the structured pseudospectrum. Here, we use \texttt{NumPy}, \texttt{SciPy}, \texttt{python-flint}, and \texttt{threadpoolctl} in \texttt{Python} to solve the optimization problem.

\section{Proof of inequalities between three distances}\label{app: inequality_between_three_d_proof}
In this appendix we prove the chain of inequalities relating the three distances. Fix any positive integer $m\ge1$ and $\omega\in\mathbb{C}$. For convenience,  we define a set given by
\begin{eqnarray}\label{F_omega_set}
    \mathcal{F}_{\omega}=\Big\{\mathbf{c}\in\mathbb{R}^m: \det \mathbf{D}(\mathbf{c})=0\Big\}\, ,
\end{eqnarray}
where $\mathbf{D}(\mathbf{c})$ can de found in Eq. (\ref{D_matrix}). It is known for us that $0\in\lambda(\omega \mathbf{I}-\mathbf{L}-\delta \mathbf{L})$ is equivalent to $\det\mathbf{D}(\mathbf{c})=0$, from which two structured spectral distances $d_{\mathrm{str},\mathcal{S}}(\omega)$ and $d_{\mathrm{str},\mathrm{E}}(\omega)$ are rewriten into
\begin{eqnarray}
    d_{\mathrm{str},\mathcal{S}}(\omega)=\inf_{\mathbf{c}\in\mathcal{F}_{\omega}}\|\widetilde{\delta \mathbf{L}}(\mathbf{c})\|_{\mathrm{HS}}\, ,\quad \text{and} \quad d_{\mathrm{str},\mathrm{E}}(\omega)=\inf_{\mathbf{c}\in\mathcal{F}_{\omega}}\|\widetilde{\delta \mathbf{L}}(\mathbf{c})\|_{2}\, .
\end{eqnarray}
From $\mathbf{D}(\mathbf{c})=\omega\mathbf{I}-\mathbf{L}-\delta \mathbf{L}(\mathbf{c})$, one has the whiten matrix $\widetilde{\mathbf{D}}(0)=\mathbf{W}\mathbf{D}(0)\mathbf{W}^{-1}=\mathbf{W}(\omega\mathbf{I}-\mathbf{L})\mathbf{W}^{-1}$. For any $\mathbf{c}\in\mathcal{F}_{\omega}$, we take a whiten unit vector $\mathbf{v}_{\text{c}}$, and get $\widetilde{\mathbf{D}}(\mathbf{c})\mathbf{v}_{\text{c}}=\Big(\widetilde{\mathbf{D}}(0)-\widetilde{\delta \mathbf{L}}(\mathbf{c})\Big)\mathbf{v}_{\text{c}}=0$. Therefore, 
\begin{eqnarray}
    d_{\rm unstr}(\omega)=\min_{\|\mathbf{v}\|_2=1}\|\widetilde{\mathbf{D}}(0)\mathbf{v}\|_2\leq\|\widetilde{\mathbf{D}}(0)\mathbf{v}_{\text{c}}\|_2=\|\widetilde{\delta \mathbf{L}}(\mathbf{c})\mathbf{v}_{\text{c}}\|_2\leq\|\widetilde{\delta \mathbf{L}}(\mathbf{c})\|_2\, .
\end{eqnarray}
Taking the infimum with respect to $\mathbf{c}\in\mathcal{F}_{\omega}$, we obtain $d_{\rm unstr}(\omega)\leq d_{\mathrm{str},\mathrm{E}}(\omega)$.

Assume that the rank of $\widetilde{\delta \mathbf{L}}$ is $r>0$ and its non-zero singular values $s_1\ge \cdots\ge s_r>0$, then one has
\begin{eqnarray}
    \|\widetilde{\delta \mathbf{L}}(\mathbf{c})\|_2^2=s_1^2\le\sum_{j=1}^{r}s_j^2=\|\widetilde{\delta \mathbf{L}}(\mathbf{c})\|_{\mathrm{HS}}^2\le rs_1^2 \le r_{\star}s_1^2\, ,
\end{eqnarray}
where $r_{\star}$ denotes the maximal rank of all possible admissible $\widetilde{\delta \mathbf{L}}$, i.e., $\operatorname{rank}$ $\widetilde{\delta \mathbf{L}}(\mathbf{c})\le r_{\star}$ is satisfied for any $\mathbf{c}\in\mathbb{R}^m$. Note that $1\leq r_{\star}\leq n_{\text{U}}$. Taking the square root of the above, we have
\begin{eqnarray}\label{inequalities_between_2_HS_2}
    \|\widetilde{\delta \mathbf{L}}(\mathbf{c})\|_2\le\|\widetilde{\delta \mathbf{L}}(\mathbf{c})\|_{\mathrm{HS}}\le \sqrt{r_{\star}}\|\widetilde{\delta \mathbf{L}}(\mathbf{c})\|_2\, .
\end{eqnarray}
From the left inequality of (\ref{inequalities_between_2_HS_2}), taking infimum over $\mathcal{F}_{\omega}$, we arrive at $d_{\mathrm{str,E}}(\omega)\le d_{\mathrm{str},\mathcal{S}}(\omega)$. For $\eta>0$, according to the definition of infimum, taking $\mathbf{c}_\eta\in\mathcal{F}_{\omega}$, one has $\|\widetilde{\delta\mathbf{L}}(\mathbf{c}_\eta)\|_2\leq d_{\mathrm{str,E}}(\omega)+\eta$. From the right inequality of (\ref{inequalities_between_2_HS_2}), we have
\begin{eqnarray}
    d_{\mathrm{str},\mathcal{S}}(\omega)\leq\|\widetilde{\delta \mathbf{L}}(\mathbf{c}_{\eta})\|_{\mathrm{HS}}\leq\sqrt{r_{\star}}\Big(d_{\mathrm{str,E}}(\omega)+\eta\Big)\, .
\end{eqnarray}
Let $\eta\to0$, $d_{\mathrm{str},\mathcal{S}}(\omega)\leq\sqrt{r_{\star}}d_{\mathrm{str,E}}(\omega)$ is derived. Since $\widetilde{\delta\mathbf{L}}=\mathbf{W}(\delta \mathbf{L})\mathbf{W}^{-1}$, the rank of $\delta \mathbf{L}$ equals to the rank of $\widetilde{\delta\mathbf{L}}$. From Eq. (\ref{S_EFT}) and Eq. (\ref{S_RZ}), it can be seen that $\delta V/p\big|_{\sigma=0}=0$ for both EFT and RZ cases. Therefore, the rank of $\delta \mathbf{L}(\mathbf{c})$ is satisfied with
\begin{eqnarray}
    \operatorname{rank}\, \delta\mathbf{L}(\mathbf{c})\leq N_{\text{c}}\, .
\end{eqnarray}
Since there exist admissible perturbations $\delta \mathbf{L}(\mathbf{c})$ of rank exactly $N_\text{c}$, so we have $r_{\star}=N_{\text{c}}$ and get (\ref{inequality_between_three_d}).

\section{Local elliptical asymptotics of the pseudospectrum contours for $d_{\mathrm{str},\mathcal{S}}(\omega)$}\label{app: elliptical_proof}
\begin{theorem}
We assume that $\omega_n$ be an isolated algebraically simple eigenvalue of the finite-dimensional matrix $\mathbf{L}$, where $n$ refers to the overtone number. Let $\mathbf{U}_n$ and $\mathbf{U}_n^\mathrm{L}$ denote the corresponding right and left eigenvectors, normalized according to
\begin{eqnarray}
    \mathbf{L}\mathbf{U}_n=\omega_n \mathbf{U}_n\, ,\quad (\mathbf{U}_n^\mathrm{L})^\dagger \mathbf{L}=\omega_n (\mathbf{U}_n^\mathrm{L})^\dagger\, ,\quad (\mathbf{U}_n^\mathrm{L})^\dagger \mathbf{U}_n=1\, .
\end{eqnarray}
Define the first-order eigenvalue sensitivities by
\begin{eqnarray}
    \alpha_{n,k}:=\mathrm{i}(\mathbf{U}_n^\mathrm{L})^\dagger \mathbf{M}_k \mathbf{U}_n\, ,\quad k=1,\cdots,m\, ,
\end{eqnarray}
and introduce the real sensitivity matrix
\begin{eqnarray}\label{def_J}
    \mathbf{J}_n:=\begin{bmatrix}
        \operatorname{Re}\alpha_{n,1}&\cdots&\operatorname{Re}\alpha_{n,m}\\
        \operatorname{Im}\alpha_{n,1}&\cdots&\operatorname{Im}\alpha_{n,m}
    \end{bmatrix}
    \in\mathbb{R}^{2\times m}\, .
\end{eqnarray}
For the offset of the original QNM spectrum $\omega_n$,
\begin{eqnarray}\label{def_z_and_x}
    z:=\omega-\omega_n\, ,\quad\mathbf{x}:=
    \begin{bmatrix}
        \operatorname{Re}z\\
        \operatorname{Im}z
    \end{bmatrix}\, ,
\end{eqnarray}
define two matrices
\begin{eqnarray}\label{def_hat_J_and_H_n}
    \widehat{\mathbf{J}}_n:=\mathbf{J}_n \mathbf{R}^{-1}\, ,\quad \mathbf{H}_n:=\widehat{\mathbf{J}}_n\widehat{\mathbf{J}}_n^{\mathrm{T}}=\mathbf{J}_n \mathbf{G}^{-1}\mathbf{J}_n^{\mathrm{T}}\, ,
\end{eqnarray}
where $\mathbf{R}$ is given by Eq. (\ref{Cholesky_decomposition_G}) and $\mathbf{G}$ is the real symmetric positive definite matrix. If
\begin{eqnarray}
    \operatorname{rank}_{\mathbb{R}}\mathbf{J}_n=2\, ,
\end{eqnarray}
then $\mathbf{H}_n$ is positive definite with dimension $2\times2$, and the structured spectral distance $d_{\mathrm{str},\mathcal{S}}(\omega)$ satisfies
\begin{eqnarray}\label{eq:structured_distance_ellipse}
    \Big[d_{\mathrm{str},\mathcal{S}}(\omega_n+z)\Big]^2=\mathbf{x}^{\mathrm{ T}}\mathbf{H}_n^{-1}\mathbf{x}+\mathcal{O}\big(\|\mathbf{x}\|_2^3\big)
\end{eqnarray}
as $\mathbf{x}\to0$. Consequently, for sufficiently small $\varepsilon>0$, the local contour $d_{\mathrm{str},\mathcal{S}}(\omega)=\varepsilon$ is asymptotically an ellipse in the complex-frequency plane, whose leading-order equation is
\begin{eqnarray}\label{eq:structured_ellipse}
    \mathbf{x}^{\mathrm{T}}\mathbf{H}_n^{-1}\mathbf{x}=\varepsilon^2\, .
\end{eqnarray}
\end{theorem}

\begin{proof}
First, since $\omega_n$ is an isolated algebraically simple eigenvalue, for sufficiently small $\mathbf{c}$, there exists a local eigenvalue branch $\omega_n(\mathbf{c})$ and a corresponding differentiable right eigenvector
$\mathbf{U}_n(\mathbf{c})$ satisfying
\begin{eqnarray}
    \Big[\mathbf{L}+\delta \mathbf{L}(\mathbf{c})\Big]\mathbf{U}_n(\mathbf{c})=\omega_n(\mathbf{c})\mathbf{U}_n(\mathbf{c})\, .
\end{eqnarray}
Differentiating this equation with respect to $c_k$ and evaluating at $\mathbf{c}=0$ gives
\begin{eqnarray}\label{eq:eigenvalue_derivative}
    \mathrm{i}\mathbf{M}_k \mathbf{U}_n+(\mathbf{L}-\omega_n \mathbf{I})\frac{\partial \mathbf{U}_n}{\partial c_k}\bigg|_{\mathbf{c}=0}=\frac{\partial\omega_n}{\partial c_k}\bigg|_{\mathbf{c}=0}\mathbf{U}_n\, .
\end{eqnarray}
Multiplying Eq. \eqref{eq:eigenvalue_derivative} from the left by $(\mathbf{U}_n^\mathrm{L})^\dagger$, the eigenvector-derivative term vanishes because
\begin{eqnarray}
    (\mathbf{U}_n^\mathrm{L})^\dagger(\mathbf{L}-\omega_n \mathbf{I})=0\, .
\end{eqnarray}
Using $(\mathbf{U}_n^\mathrm{L})^\dagger \mathbf{U}_n=1$, we obtain
\begin{eqnarray}
    \frac{\partial\omega_n}{\partial c_k}\bigg|_{\mathbf{c}=0}=\mathrm{i}(\mathbf{U}_n^\mathrm{L})^\dagger \mathbf{M}_k \mathbf{U}_n=\alpha_{n,k}\, .
\end{eqnarray}
It follows from the Taylor expansion of the local eigenvalue branch that
\begin{eqnarray}\label{eq:frequency_taylor}
    \omega_n(\mathbf{c})-\omega_n=\sum_{k=1}^{m}\alpha_{n,k}c_k+\mathcal{O}\big(\|\mathbf{c}\|_2^2\big)\, .
\end{eqnarray}
Define the real-valued frequency-shift map
\begin{eqnarray}
    \mathbf{f}_n(\mathbf{c}):=
    \begin{bmatrix}
        \operatorname{Re}\big(\omega_n(\mathbf{c})-\omega_n\big)\\
        \operatorname{Im}\big(\omega_n(\mathbf{c})-\omega_n\big)
    \end{bmatrix}\, .
\end{eqnarray}
Eq. \eqref{eq:frequency_taylor} and Eq. (\ref{def_J}) then yield
\begin{eqnarray}\label{eq:real_frequency_taylor}
    \mathbf{f}_n(\mathbf{c})=\mathbf{J}_n\mathbf{c}+\mathcal{O}\big(\|\mathbf{c}\|_2^2\big)\, .
\end{eqnarray}
Thus, to leading order, producing the prescribed frequency shift $z$ amounts to solving
\begin{eqnarray}\label{eq:linear_frequency_constraint}
    \mathbf{J}_n\mathbf{c}=\mathbf{x}\, ,
\end{eqnarray}
where Eq. (\ref{def_z_and_x}) is used.

Second, we consider the linearized structured distance
\begin{eqnarray}\label{eq:linearized_distance}
    \rho_{n,\mathcal{S}}(\mathbf{x}):=\min_{\mathbf{J}_n\mathbf{c}=\mathbf{x}}\sqrt{\mathbf{c}^{\mathrm{T}}\mathbf{G}\mathbf{c}}\, .
\end{eqnarray}
From Eq. (\ref{Cholesky_decomposition_G}) and Eq. (\ref{def_hat_J_and_H_n}), Eq. \eqref{eq:linearized_distance} becomes
\begin{eqnarray}\label{eq:minimum_norm_problem}
    \rho_{n,\mathcal{S}}(\mathbf{x})
    =
    \min_{\widehat{\mathbf{J}}_n\mathbf{y}
    =\mathbf{x}}
    \|\mathbf{y}\|_2\, .
\end{eqnarray}
Because $\operatorname{rank}_{\mathbb{R}}\mathbf{J}_n=2$ and $\mathbf{R}$ is invertible, we have $\operatorname{rank}_{\mathbb{R}} \widehat{\mathbf{J}}_n=2$. Therefore, for every nonzero $\mathbf{u}\in\mathbb{R}^2$,
\begin{eqnarray}
    \mathbf{u}^{\mathrm{T}}\mathbf{H}_n\mathbf{u}=\|\widehat{\mathbf{J}}_n^{\mathrm{T}}\mathbf{u}\|_2^2>0\, .
\end{eqnarray}
Hence $\mathbf{H}_n$ is positive definite and invertible. Define
\begin{eqnarray}\label{eq:minimum_norm_solution}
    \mathbf{y}_0:=\widehat{\mathbf{J}}_n^{\mathrm{T}}\mathbf{H}_n^{-1}\mathbf{x}\, .
\end{eqnarray}
By construction, we have
\begin{eqnarray}
    \widehat{\mathbf{J}}_n\mathbf{y}_0=
    \mathbf{H}_n\mathbf{H}_n^{-1}\mathbf{x}
    =\mathbf{x}\, ,
\end{eqnarray}
so $\mathbf{y}_0$ is feasible for Eq. \eqref{eq:minimum_norm_problem}. Every other feasible vector can be written as $\mathbf{y}_0+\mathbf{h}$, where $\widehat{\mathbf{J}}_n\mathbf{h}=0$. Moreover,
\begin{eqnarray}
    \mathbf{y}_0^{\mathrm{T}}\mathbf{h}=\mathbf{x}^{\mathrm{T}}\mathbf{H}_n^{-1}\widehat{\mathbf{J}}_n\mathbf{h}=0\, .
\end{eqnarray}
Thus $\mathbf{y}_0$ is orthogonal to the null space of $\widehat{\mathbf{J}}_n$, and
\begin{eqnarray}
    \|\mathbf{y}_0+\mathbf{h}\|_2^2=\|\mathbf{y}_0\|_2^2+\|\mathbf{h}\|_2^2\, .
\end{eqnarray}
The minimum is therefore uniquely attained at
$\mathbf{y}=\mathbf{y}_0$. It follows that
\begin{eqnarray}\label{eq:linearized_elliptical_distance}
    \Big[\rho_{n,\mathcal{S}}(\mathbf{x})\Big]^2=\|\mathbf{y}_0\|_2^2=\mathbf{x}^{\mathrm{T}}\mathbf{H}_n^{-1}\widehat{\mathbf{J}}_n\widehat{\mathbf{J}}_n^{\mathrm{T}}\mathbf{H}_n^{-1}\mathbf{x}=\mathbf{x}^{\mathrm{T}}\mathbf{H}_n^{-1}\mathbf{x}\, .
\end{eqnarray}
This establishes the exact elliptical form for the linearized problem.

Third, we show that nonlinear corrections do not change the leading-order ellipse. Let
\begin{eqnarray}
    \mathbf{c}_0=\mathbf{R}^{-1}\mathbf{y}_0\, .
\end{eqnarray}
Since $\mathbf{y}_0$ depends linearly on $\mathbf{x}$, we have
\begin{eqnarray}
    \mathbf{c}_0=\mathcal{O}\big(\|\mathbf{x}\|_2\big)\, ,\qquad \mathbf{J}_n\mathbf{c}_0=\mathbf{x}\, .
\end{eqnarray}
By Eq. \eqref{eq:real_frequency_taylor}, we get
\begin{eqnarray}\label{eq:nonlinear_residual}
    \mathbf{f}_n(\mathbf{c}_0)-\mathbf{x}=\mathcal{O}\big(\|\mathbf{x}\|_2^2\big)\, .
\end{eqnarray}
To correct this residual, introduce the right inverse
\begin{eqnarray}
    \mathbf{J}_n^+:=\mathbf{J}_n^{\mathrm{T}}\bigl(\mathbf{J}_n\mathbf{J}_n^{\mathrm{T}}\bigr)^{-1}\, ,\qquad \mathbf{J}_n\mathbf{J}_n^+=\mathbf{I}_2\, ,
\end{eqnarray}
where $\mathbf{I}_2$ the $2$-dimensional  identity matrix. We then consider the equation ($\mathbf{c}_0$ is a function of $\mathbf{x}$.)
\begin{eqnarray}\label{eq:nonlinear_correction}
    \mathbf{f}_n\big(\mathbf{c}_0+\mathbf{J}_n^+\boldsymbol{\eta}\big)=\mathbf{x}\, ,\qquad\boldsymbol{\eta}\in\mathbb{R}^2\, .
\end{eqnarray}
At $\mathbf{x}=0$ and $\boldsymbol{\eta}=0$, the derivative of the left-hand side with respect to $\boldsymbol{\eta}$ is
\begin{eqnarray}
    \mathbf{J}_n\mathbf{J}_n^+=\mathbf{I}_2\, .
\end{eqnarray}
The implicit function theorem, together with Eq. \eqref{eq:nonlinear_residual}, therefore guarantees a local solution satisfying
\begin{eqnarray}
    \boldsymbol{\eta}=\mathcal{O}\big(\|\mathbf{x}\|_2^2\big)\, .
\end{eqnarray}
Consequently, there exists an exactly feasible perturbation
\begin{eqnarray}
    \mathbf{c}_{\mathrm{corr}}=\mathbf{c}_0+\mathbf{J}_n^+\boldsymbol{\eta}=\mathbf{c}_0+\mathcal{O}\big(\|\mathbf{x}\|_2^2\big)\, ,
\end{eqnarray}
which produces the prescribed frequency shift on the local eigenvalue branch. Using the norm induced by $\mathbf{G}$, we obtain
\begin{eqnarray}
    \|\mathbf{c}_{\mathrm{corr}}\|_{\mathrm{G}}
    =
    \rho_{n,\mathcal{S}}(\mathbf{x})
    +\mathcal{O}\big(\|\mathbf{x}\|_2^2\big)\, ,\quad \text{with} \quad \|\mathbf{c}_{\mathrm{corr}}\|_{\mathrm{G}}=\sqrt{\mathbf{c}_{\mathrm{corr}}^{\mathrm{T}}\mathbf{G}\mathbf{c}_{\mathrm{corr}}}\, ,\quad \text{and}\quad \rho_{n,\mathcal{S}}(\mathbf{x})=\|\mathbf{c}_{0}\|_{\mathrm{G}}\, .
\end{eqnarray}
Thus the exact structured distance $d_{\mathrm{str},\mathcal{S}}(\omega)$ satisfies the upper bound
\begin{eqnarray}\label{eq:distance_upper_bound}
    d_{\mathrm{str},\mathcal{S}}(\omega_n+z)\leq\rho_{n,\mathcal{S}}(\mathbf{x})+\mathcal{O}\big(\|\mathbf{x}\|_2^2\big)\, .
\end{eqnarray}

For the reverse bound, let $\mathbf{c}_{\star}$ be an exact minimizing perturbation. Its existence follows from the finite-dimensional constrained minimization problem and the positive definiteness of $\mathbf{G}$. From Eq. \eqref{eq:distance_upper_bound}, we have\footnote{From Eq. (\ref{eq:linearized_elliptical_distance}), $\rho_{n,\mathcal{S}}(\mathbf{x})=\mathcal{O}\big(\|\mathbf{x}\|_2\big)$. For $\|\mathbf{x}\|_2\le 1$, there exists $C_1>0$, $C_2>0$ such that $d_{\mathrm{str},\mathcal{S}}(\omega_n+z)\leq C_1\|\mathbf{x}\|_2+C_2\|\mathbf{x}\|_2^2\le\max\{C_1,C_2\}\|\mathbf{x}\|_2$.}
\begin{eqnarray}
    d_{\mathrm{str},\mathcal{S}}(\omega_n+z)=\mathcal{O}\big(\|\mathbf{x}\|_2\big)\, .
\end{eqnarray}
Hence
\begin{eqnarray}
    \lambda_{\min}(\mathbf{G})\|\mathbf{c}_{\star}\|_2^2\leq\mathbf{c}_{\star}^{\mathrm{T}}\mathbf{G}\mathbf{c}_{\star}=\Big[d_{\mathrm{str},\mathcal{S}}(\omega_n+z)\Big]^2=\mathcal{O}\big(\|\mathbf{x}\|_2^2\big)\, ,
\end{eqnarray}
which implies
\begin{eqnarray}\label{c_star}
    \mathbf{c}_{\star}=\mathcal{O}\big(\|\mathbf{x}\|_2\big)\, .
\end{eqnarray}
Since $\mathbf{c}_{\star}$ is feasible on the local eigenvalue branch, $\mathbf{f}_n(\mathbf{c}_{\star})=\mathbf{x}$. Therefore, Eq. \eqref{eq:real_frequency_taylor} and Eq. (\ref{c_star}) give
\begin{eqnarray}
    \mathbf{x}-\mathbf{J}_n\mathbf{c}_{\star}=\mathcal{O}\big(\|\mathbf{x}\|_2^2\big)\, .
\end{eqnarray}
The function $\rho_{n,\mathcal{S}}$ is a norm on $\mathbb{R}^2$, as follows from Eq. \eqref{eq:linearized_elliptical_distance} and $\mathbf{H}_n$ is positive definite. By its definition and the triangle inequality, one has the follow inequalities 
\begin{eqnarray}\label{eq:distance_lower_bound}
    \rho_{n,\mathcal{S}}(\mathbf{x})\leq\rho_{n,\mathcal{S}}(\mathbf{J}_n\mathbf{c}_{\star})+\rho_{n,\mathcal{S}}(\mathbf{x}-\mathbf{J}_n\mathbf{c}_{\star})\leq\|\mathbf{c}_{\star}\|_{\mathrm{G}}+\mathcal{O}\big(\|\mathbf{x}\|_2^2\big)=d_{\mathrm{str},\mathcal{S}}(\omega_n+z)+\mathcal{O}\big(\|\mathbf{x}\|_2^2\big)\, .
\end{eqnarray}
For the second inequality, we can obtain it following way. From $\rho_{n,\mathcal{S}}(\mathbf{J}_n\mathbf{c}_{\star})=\min_{\mathbf{J}_n\mathbf{c}=\mathbf{J}_n\mathbf{c}_{\star}}\sqrt{\mathbf{c}^{\mathrm{T}}\mathbf{G}\mathbf{c}}$, we have $\rho_{n,\mathcal{S}}(\mathbf{J}_n\mathbf{c}_{\star})\leq\|\mathbf{c}_{\star}\|_{\mathrm{G}}$, because $\mathbf{c}=\mathbf{c}_{\star}$ is a feasible solution. In addition, for $\rho_{n,\mathcal{S}}(\mathbf{x}-\mathbf{J}_n\mathbf{c}_{\star})$, from Eq. (\ref{eq:linearized_elliptical_distance}), we have
\begin{eqnarray}
    \rho_{n,\mathcal{S}}(\mathbf{x}-\mathbf{J}_n\mathbf{c}_{\star})=\sqrt{(\mathbf{x}-\mathbf{J}_n\mathbf{c}_{\star})^{\mathrm{T}}\mathbf{H}_n^{-1}(\mathbf{x}-\mathbf{J}_n\mathbf{c}_{\star})}\leq \sqrt{\|\mathbf{H}_n^{-1}\|_2}\cdot\|\mathbf{x}-\mathbf{J}_n\mathbf{c}_{\star}\|_2=\mathcal{O}\big(\|\mathbf{x}\|_2^2\big)\, .
\end{eqnarray}
Combining Eqs. \eqref{eq:distance_upper_bound} and
\eqref{eq:distance_lower_bound}, we conclude that
\begin{eqnarray}\label{diff_d_s_and_rho}
    d_{\mathrm{str},\mathcal{S}}(\omega_n+z)=\rho_{n,\mathcal{S}}(\mathbf{x})+\mathcal{O}\big(\|\mathbf{x}\|_2^2\big)\, .
\end{eqnarray}
This tells us that the difference between $d_{\mathrm{str},\mathcal{S}}(\omega_n+z)$ and $\rho_{n,\mathcal{S}}(\mathbf{x})$ is $\mathcal{O}\big(\|\mathbf{x}\|_2^2\big)$ as $\mathbf{x}\to0$. Both distances $d_{\mathrm{str},\mathcal{S}}(\omega_n+z)$ and $\rho_{n,\mathcal{S}}(\mathbf{x})$ are $\mathcal{O}\big(\|\mathbf{x}\|_2\big)$, and hence one obtains
\begin{eqnarray}
    \Big[d_{\mathrm{str},S}(\omega_n+z)\Big]^2-\Big[\rho_{n,\mathcal{S}}(\boldsymbol{x})\Big]^2=\Big[d_{\mathrm{str},\mathcal{S}}(\omega_n+z)-\rho_{n,\mathcal{S}}(\mathbf{x})\Big]\Big[d_{\mathrm{str},\mathcal{S}}(\omega_n+z)+\rho_{n,\mathcal{S}}(\mathbf{x})\Big]=\mathcal{O}\big(\|\mathbf{x}\|_2^3\big)\, ,
\end{eqnarray}
where Eq. (\ref{diff_d_s_and_rho}) is used. Substituting Eq. \eqref{eq:linearized_elliptical_distance} establishes
\begin{eqnarray}
    \Big[d_{\mathrm{str},S}(\omega_n+z)\Big]^2=\mathbf{x}^{\mathrm{T}}\mathbf{H}_n^{-1}\mathbf{x}+\mathcal{O}\big(\|\mathbf{x}\|_2^3\big)\, ,
\end{eqnarray}
which proves the asserted asymptotic formula.

Finally, since $\mathbf{H}_n$ is real symmetric and positive definite, there exists a real orthogonal matrix $\mathbf{Q}_n$ such that
\begin{eqnarray}
    \mathbf{H}_n=\mathbf{Q}_n
    \begin{bmatrix}
        s_{n,1}^2&0\\
        0&s_{n,2}^2
    \end{bmatrix}
    \mathbf{Q}_n^{\mathrm{T}}\, ,\qquad s_{n,1}\geq s_{n,2}>0\, ,
\end{eqnarray}
where $s_{n,1}^2$ and $s_{n,2}^2$ are the eigenvalues of $\mathbf{H}_n$. Introduce the principal-axis coordinates
\begin{eqnarray}
    \boldsymbol{\xi}=\mathbf{Q}_n^{\mathrm{T}}\mathbf{x}\, .
\end{eqnarray}
The leading-order contour $d_{\mathrm{str},\mathcal{S}}(\omega_n+z)=\varepsilon$ then becomes a standard elliptical trajectory equation as follow
\begin{eqnarray}\label{standard_elliptical_trajectory_equation}
    \frac{\xi_1^2}{(\varepsilon s_{n,1})^2}+\frac{\xi_2^2}{(\varepsilon s_{n,2})^2}=1\, .
\end{eqnarray}
Thus, the semiaxes are $\varepsilon s_{n,1}$ and $\varepsilon s_{n,2}$, and their directions are determined by the columns of $\mathbf{Q}_n$.

On the exact contour, $\|\mathbf{x}\|_2=\mathcal{O}(\varepsilon)$. The asymptotic distance formula therefore implies
\begin{eqnarray}
    \frac{\mathbf{x}^{\mathrm{T}}\mathbf{H}_n^{-1}\mathbf{x}}{\varepsilon^2}=1+\mathcal{O}(\varepsilon)\, .
\end{eqnarray}
Hence, as $\varepsilon\to0$, the appropriately rescaled exact contours converge to the ellipse determined by $\mathbf{H}_n$. Higher-order terms produce deviations from this ellipse at finite perturbation $\varepsilon$. This completes the proof.
\end{proof}

\bibliography{reference}
\bibliographystyle{apsrev4-1}

\end{document}